\documentclass[twocolumn,usenatbib]{mnras}
\usepackage{graphicx}
\usepackage{amsmath}
\usepackage{amssymb}
\usepackage{physics}
\usepackage{xfrac}
\usepackage{comment}
\usepackage{hyperref}
\MakeRobust{\hyperlink}
\usepackage{url}
\usepackage{xcolor}

\newcommand{\dex}{\,\mathrm{dex}}
\newcommand{\msun}{\ensuremath{\text{M}_\odot}}

\usepackage{array}
\newcolumntype{C}[1]{>{\centering\let\newline\\\arraybackslash\hspace{0pt}}m{#1}}

\title[Mass profiles of dwarf galaxies]{The mass profiles of dwarf galaxies from Dark Energy Survey lensing}

\author[J. Thornton, A. Amon et al.]{Joseph Thornton$^{1}$\thanks{E-mail: jdt50@cantab.ac.uk}, 
Alexandra Amon$^{2,1,3}$\thanks{E-mail: alexandra.amon@princeton.edu},
Risa H. Wechsler$^{4,5}$\thanks{E-mail: rwechsler@stanford.edu},
Susmita Adhikari$^{6}$,
Yao-Yuan Mao$^{7}$, 
\newauthor
Justin Myles$^{2,4}$, 
Marla Geha$^{8}$,
Nitya Kallivayalil$^{9}$,
Erik Tollerud$^{10}$,
Benjamin Weiner$^{11}$\\
${1}$ Institute of Astronomy, Madingley Road, Cambridge, CB3 OHA, United Kingdom \\
${2}$ Department of Astrophysical Sciences, Princeton University, Princeton, NJ 08544, USA \\  
${3}$ Kavli Institute for Cosmology Cambridge, Madingley Road, Cambridge, CB3 OHA, United Kingdom \\
${4}$ Kavli Institute for Particle Astrophysics and Cosmology and Department of Physics, Stanford University, Stanford, CA 94305, USA \\
${5}$ SLAC National Accelerator Laboratory, Menlo Park, CA 94025, USA\\
${6}$ Department of Physics, Indian Institute of Science Education and Research, Homi Bhaba Road, Pashan, Pune 411008, India\\
${7}$ Department of Physics and Astronomy, University of Utah, Salt Lake City, UT 84112, USA\\
${8}$ Department of Astronomy, Yale University, New Haven, CT 06520, USA\\
${9}$ Department of Astronomy, University of Virginia, 530 McCormick Road, Charlottesville, VA 22904, USA
\\
${10}$ Space Telescope Science Institute, 3700 San Martin Dr., Baltimore, MD 21218, USA
\\
${11}$ MMT and Steward Observatories, University of Arizona, 933 North Cherry Avenue, Tucson, AZ 85721-0065, USA
} 

\date{Accepted XXX. Received YYY; in original form ZZZ}

\pubyear{2023}

\begin{document}

\thinmuskip=1mu
\medmuskip=1mu
\thickmuskip=3mu

\maketitle

\begin{abstract} 
We present a novel approach to extracting dwarf galaxies from photometric data to measure their average halo mass profile with weak lensing. We characterise their stellar mass and redshift distributions with a spectroscopic calibration sample.
By combining the $\sim5000\mathrm{deg}^2$ multi-band photometry from the Dark Energy Survey and redshifts from the Satellites Around Galactic Analogs (SAGA) survey with an unsupervised machine learning method, we select a low-mass galaxy sample spanning redshifts $z<0.3$ and divide it into three mass bins.
From low to high median mass, the bins contain [146 420, 330 146, 275 028] galaxies and have median stellar masses of $\log_{10}(M_*/\msun)=\left[8.52\substack{+0.57 \\ -0.76},\,9.02\substack{+0.50 \\ -0.64},\,9.49\substack{+0.50 \\ -0.58}\right]$\,.
We measure the stacked excess surface mass density profiles, $\Delta\Sigma(R)$, of these galaxies using galaxy--galaxy lensing with a signal-to-noise of [14, 23, 28].
Through a simulation-based forward-modelling approach, we fit the measurements to constrain the stellar-to-halo mass relation and find the median halo mass of these samples to be $\log_{10}(M_{\rm halo}/\msun)$ = [\,$10.67\substack{+0.2\\-0.4}$,\,$11.01\substack{+0.14 \\ -0.27}$,\,$11.40\substack{+0.08\\-0.15}$\,]. The CDM profiles are consistent with NFW profiles over scales $\lesssim 0.15 \,\,\rm{h}^{-1}$ Mpc.
We find that $\sim20$ per cent of the dwarf galaxy sample are satellites.
This is the first measurement of the halo profiles and masses of such a comprehensive, low-mass galaxy sample. The techniques presented here pave the way for extracting and analysing even lower-mass dwarf galaxies and for more finely splitting galaxies by their properties with future photometric and spectroscopic survey data.

\end{abstract}


\section{Introduction}
The success of the Lambda Cold Dark Matter ($\Lambda$CDM) cosmological model has been reinforced by percent-level constraints on its cosmological parameters using a number of observations \citep[e.g.,][]{planck18,Alam21}.
This CDM-dominated framework is the foundation for a universe where structure forms hierarchically, with the growth of large dark matter haloes and their residing galaxies arising from accretion and mergers of smaller ones \citep{Peebles1980}.  
The formation and evolution of galaxies is tied to the growth of their dark matter haloes, and is explored as the `galaxy–-halo connection' \citep{Wechsler_2018}.
Indeed, the characteristics of these haloes can be used to predict the properties of the galaxies within them \citep[e.g.,][]{Behroozi2013}.
While large-scale cosmological measurements are well-described by this model, the nature of dark matter remains elusive, and observations from small scales ($\sim1$Mpc) reveal a less-complete picture of the cosmological model \citep[see, for a review][]{BullockBoylan2017}.
Characterising dark matter at these scales could distinguish between $\Lambda$CDM and alternative models that modify small-scale cosmic structure, however, this demands an accurate understanding of galaxy formation physics. 

Dwarf galaxies
provide a unique probe of the nature of dark matter on small scales and our understanding of the galaxy--halo connection 
\citep{deBlok10, BullockBoylan2017, CollinsRead2022}. 
In this work, we focus on bright dwarf galaxies, defined by a stellar mass range of $M_*\approx10^{7-9.5} \msun$.
Historically, observations of Milky Way dwarf galaxies, thought to be dark matter dominated, have tested our understanding of the nature of dark matter and its interplay with galaxy formation \citep{deBlok:1997zlw, vandenBosch:1999ka, Klypin:1999uc, 2011MNRAS.415L..40B}. Even with an improved understanding of feedback processes in low-mass galaxies, and higher-resolution simulations\citep[e.g.,][]{BrooksZolotov14, Wetzel16, Brooks17, Martin-Alvarez2022}, some of these challenges remain. Importantly, the inner regions of dwarf galaxies’ halo profiles are highly sensitive to different dark matter models and feedback physics \citep[e.g.,][]{Tollet2016}
Non-standard dark matter theories have been proposed as potential alternatives to explain the properties of dwarf galaxies \citep{Fry2015, Kamada2017, Robles2017, Buckley:2017ijx, Tulin:2017ara, Fitts2019, Nadler2021, adhikari2022sidm},
but direct observations of dwarf galaxies' mass profiles are required to distinguish among different dark matter models. 

Measurements of the mass dependence of observed dwarf properties are also important for understanding galaxy formation physics, enabling the disentanglement of baryonic effects from dark matter properties \citep{Nadler2020,Munshi2021, Danieli2022}.
Indeed, the stellar-to-halo mass relation (SHMR), which measures the ratio of stellar mass and mass from a dark matter halo across the spectrum of halo masses, is poorly constrained at halo masses below $\sim 10^{12}\msun$. In this low-mass regime, winds from massive stars reduce the star formation efficiency with decreasing mass producing a trend of roughly $M_*\sim M_{\rm halo}^{2-3}$ \citep{Wechsler_2018}. 
Constraining the normalisation, slope, and scatter in the galaxy--halo connection in this regime could provide critical tests of the galaxy--halo connection and CDM. A collated sample of faint, distant, dwarf galaxies and precise measurements of their halo mass profiles would create a test-bed for physics that impacts small-scale cosmic structure. 

Traditionally, dark matter haloes of dwarf galaxies have been measured using kinematics \citep{Simon2019}, including rotation curves \citep[e.g.,][]{Rubin1978, Ferrero2012} of galaxy stars and gas, which typically probe only the inner regions and are typically for near-field galaxies. Alternatively, they have been inferred from models that rely on theoretical assumptions rather than direct measurements.
Examples of the latter include empirical modelling \citep[e.g.,][]{Behroozi2019},
 halo occupation distribution modelling \citep[e.g.,][]{Yang2012}, or abundance matching \citep[e.g.,][]{ConroyWechsler2009, Behroozi2013, Moster2018}.
 Weak gravitational lensing is complementary to these approaches \citep{Brainerd1996, BartelmannSchneider2001,HoekstraJain2008}. `Galaxy--galaxy lensing' (GGL) correlates the average weak lensing distortion from background `source' galaxies with the positions of a sample of foreground `lens' galaxies.
This method holds immense potential because it measures the full halo mass profile, including the dark matter and baryonic content over a large dynamic range inside and outside the halo, without assumptions about the shape or properties of the halo. GGL has proven to be an effective technique to measure the total matter profile of galaxies, typically at higher redshifts ($z>0.2$), and for halo masses above $M_{\rm halo} > 10^{11} \msun$ \citep{Mandelbaum2006,Leauthaud2012,Velander2014,Han2014,Hudson2014,Mandelbaum2015,van_Uitert_2016,Dvornik2020,Wang2021,DES:2021olg, Voice2022,Chaurasiya:2023azn}. Weak lensing profiles of large statistical low-mass samples give us a handle on low-mass galaxy physics in a cosmological context.

Extending GGL to a low-redshift ($z<0.2$), low-mass ($M_{\rm halo} < 10^{11} \msun$), regime is challenging due to the need for a large and pure dwarf galaxy sample as lenses. Selecting a sample of faint, low-redshift, and low-mass objects from large, deep photometric data requires effective methods to distinguish the low-redshift galaxy targets from a dominant background of high-redshift objects. 
While this challenge can be mitigated by spectroscopic surveys, existing spectroscopic resources are still too scarce to allow us to make use of all the available photometric data. 
Narrow-band imaging surveys, such as the Merian Survey \citep{2305.19310}, have also been proposed to tackle this challenge. There also exist several methods that use machine learning algorithms to select or estimate redshift information for low-redshift galaxies based on photometric data \citep[e.g.,][]{Tanoglidis, Wu2022, Dey2022}.
Nevertheless, the solution used in this work is to adopt the state-of-the-art approach of weak lensing redshift calibration, which involves calibrating the redshift distribution of large photometric data using spectroscopic samples that are much smaller, and span less area. The key requirement is that the spectroscopic sample must be representative of the photometry, in order to understand the colour--redshift relation \citep[see][for a review]{2022ARA&A..60..363N}. 

In this work, we combine a relatively small, but information-rich, spectroscopic dataset from the Satellites Around Galactic Analogs Survey \citep[SAGA;][]{Geha_2017, Mao_2021}, a wide-area photometric catalogue from the Dark Energy Survey \citep[DES;][]{Abbott_2021}, and the Self-organising Map algorithm \citep[SOM;][]{Kohonen1982, kohonen1990} to perform GGL analysis of a dwarf galaxy sample for the first time. 
We build a large low-mass sample over the full 5000deg$^2$ DES footprint, using a photometric selection developed by the SAGA Survey and optimised for low-redshift ($z \sim 0.01$) science \citep{Mao_2021,Elise2023}. 
We then use a SOM to sort the photometry and characterise it with the spectroscopic redshifts and stellar masses obtained from SAGA, which allows us to extract samples of low-mass galaxies in multiple stellar mass bins. We measure the stacked GGL profile of three samples with different average stellar masses and constrain the average halo mass and density profiles.
Our results also place the first direct constraints on the relationship between a galaxy's stellar mass and that of its dark matter halo in this mass regime.

In Section~\ref{sec:data}, we introduce the data used in this analysis to create the source and lens samples.
Section~\ref{sec:som} outlines the galaxy sample selection and characterisation using a self-organising map. 
Section~\ref{sec:ggl} details the GGL measurement and Section~\ref{sec:Modelling} describes our modelling of the halo mass profiles to constrain the SHMR.
Finally, in Section~\ref{sec:disc} we discuss the results and in Section~\ref{sec:summary} we summarise our findings and discuss the various methods used and their importance for the future.

\section{Data}\label{sec:data}

\begin{figure*}
    \centering
    \includegraphics[width=\textwidth]{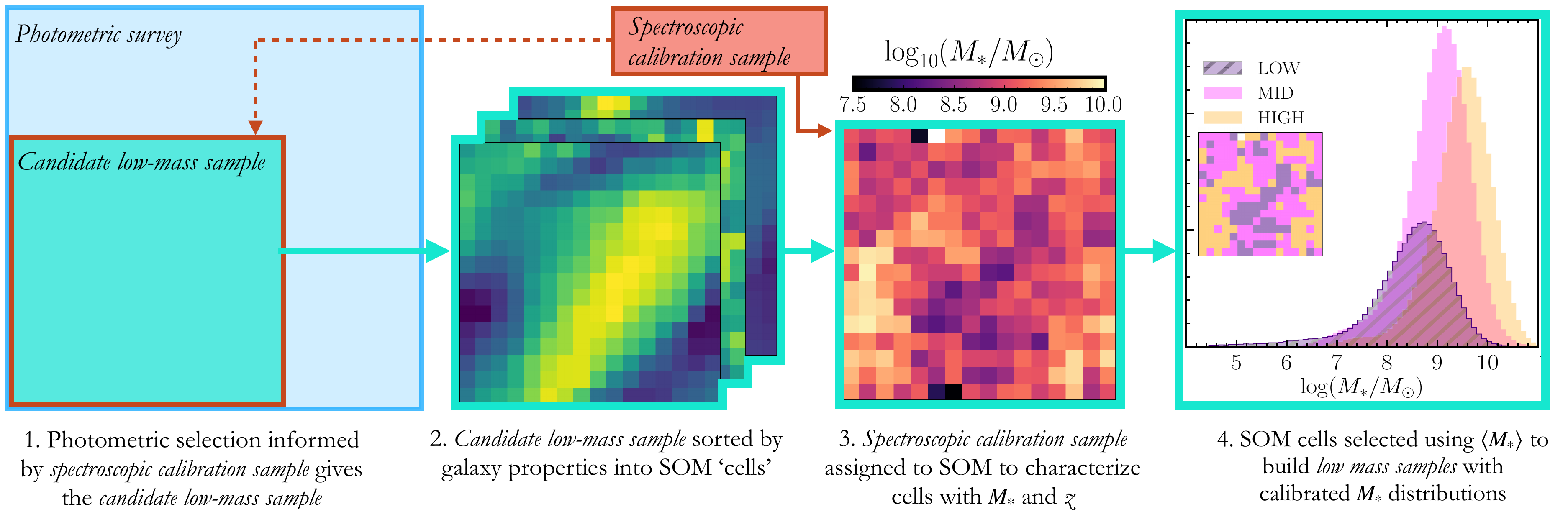}
    \caption{Schematic illustrating the methodology for selecting and characterising a \textit{low-mass sample} from a photometric survey using a \textit{spectroscopic calibration sample} and a machine learning technique. The \textit{low-mass sample} serves as the `lenses' for the GGL measurements.}
    \label{fig:scheme}
\end{figure*}
   
\begin{table*}
    \centering
    \caption{Summary of the four main galaxy samples used and produced in this work.}
    \def\arraystretch{1.1}%
    \begin{tabular}{cp{115mm}c}
    \hline
    Sample name & Data description\\
    \hline
     \textit{Spectroscopic calibration sample} & Compiled by the SAGA team, this sample supplies the spectroscopic redshifts and stellar masses used to characterise the \textit{candidate low-mass sample} in the SOM.  \\
    \textit{Candidate low-mass sample} & Selected from the DES DR2 photometry with criteria informed by the \textit{spectroscopic calibration sample}. \\
    \textit{Low-mass samples: } \textsc{LOW, MID, HIGH} & Subsets of the \textit{candidate low-mass sample} selected using the SOM by their $M_*$, to be used as the GGL lens samples.\\
   \textit{Source sample} & Background lensing photometry from DES Y3. \\
    \hline
    \end{tabular}
    \label{tab:data_breakdown}
\end{table*}

We present a novel methodology to extract a large photometric \textit{low-mass sample} of galaxies with well-understood property distributions using a relatively small \textit{spectroscopic calibration sample} of confirmed low-mass galaxies. This scheme is illustrated in Figure~\ref{fig:scheme}. First, a \textit{candidate low-mass sample} is created by applying a set of photometric selection criteria based upon the \textit{spectroscopic calibration sample}. This \textit{candidate low-mass sample} is used to `train' a SOM, thereby sorting the photometric galaxies into `cells' according to their observed properties. The \textit{spectroscopic calibration sample} is also sorted into the same SOM, thus characterising each cell with a stellar mass and redshift. SOM cells are selected according to their average stellar mass, allowing us to build \textit{low-mass samples} with calibrated stellar mass probability distributions. Finally, a background \textit{source galaxy sample} is used to measure the weak lensing profiles of the \textit{low-mass samples}.
In Table~\ref{tab:data_breakdown}, we specify each of the samples used or built in this analysis.

\subsection{Spectroscopic calibration sample from the SAGA Survey}\label{sec:saga}

\begin{figure*}
    \centering
    \includegraphics[width=\textwidth]{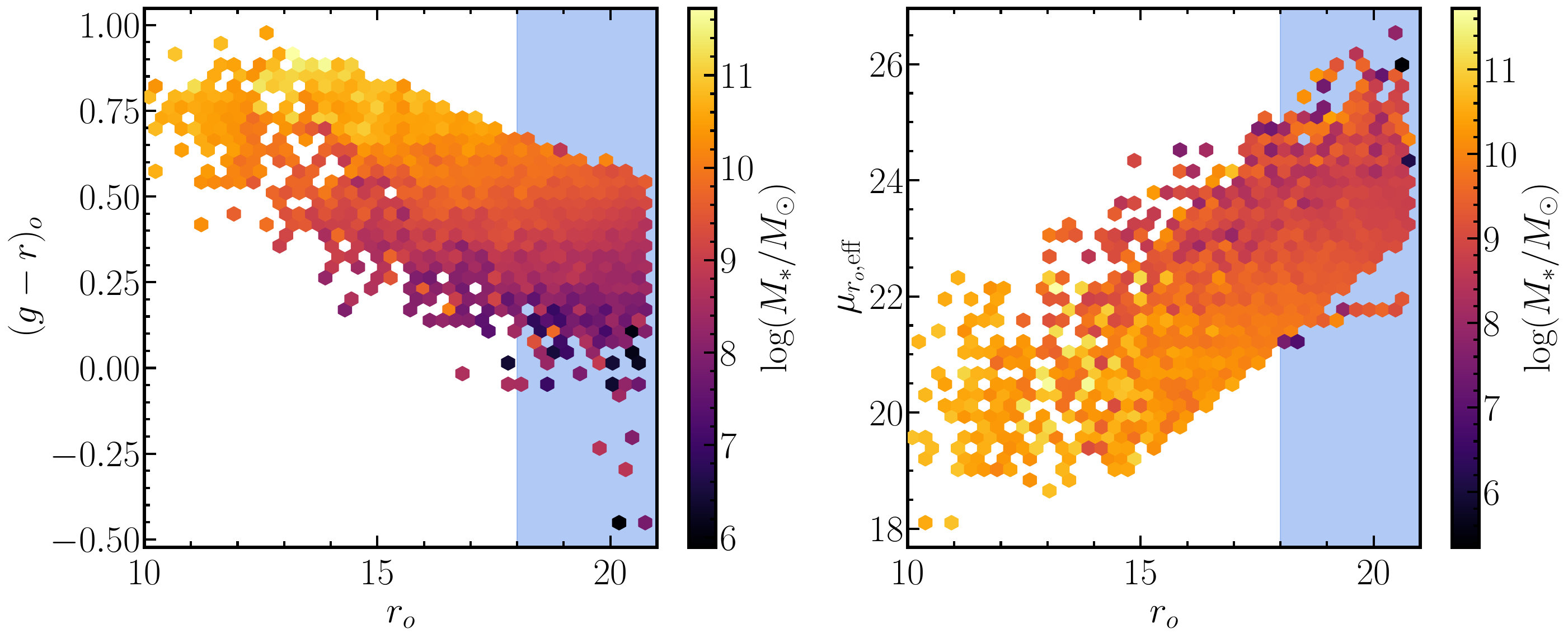}
    \caption{Distribution of the \textit{spectroscopic calibration sample} coloured by average stellar mass in a given hex bin, prior to the $r_o > 18$ selection (9,423 galaxies). The two panels show the colour--magnitude, $(g-r)_o$--$r_o$, plane (left) and surface magnitude--magnitude, $\mu_{r_o, \rm{eff}}$--$r_o$, plane (right) and are specified by the selection cuts of Eqs.~\eqref{eq:colour_cut}~and~\eqref{eq:sb_cut}. We see trends of stellar mass with these three properties, which motivates their use in the SOM training. Also, as stellar mass tends to decrease towards the fainter magnitudes, we apply a further selection to retain $r_o > 18$, highlighted in the blue region of both panels. The final calibration sample has 6,330 galaxies, with a mean stellar mass of $\langle\log_{10}(M_*/\msun)\rangle=9.13$.}
    \label{fig:saga_hexbins}
\end{figure*}

To inform the low-mass lens selection we require a \textit{spectroscopic calibration sample} that is composed of primarily low-redshift, low-mass galaxies, with basic photometric properties ($g$- and $r$-band magnitudes and $r$-band effective surface brightness) as well as spectroscopic redshifts and estimated stellar masses. 
In order to include galaxies that have masses of $10^7$ to $10^9\msun$, we construct this sample using data from the Satellites Around Galactic Analogs Survey \citep[SAGA;][]{Geha_2017, Mao_2021}; in this mass regime, SAGA includes a significant fraction of existing redshifts in the pre-DESI era.

While the SAGA Survey's primary goal is to characterise the population of satellite galaxies around 100 nearby ($z < 0.015$) Milky Way analogues, the SAGA Survey team has collected a large number of spectra for galaxies that are in the SAGA Survey footprint but with magnitudes extending to $r_o = 20.7$, fainter than the limits that most existing large spectroscopic surveys have reached. The majority of these spectra do not belong to satellite galaxies in $z < 0.015$ Milky Way systems, but to galaxies that are slightly further away, yet still have low redshift \citep[$z < 0.2$;][]{Mao_2021}.

To construct our \textit{spectroscopic calibration sample}, we choose a subset of the galaxy spectra from the SAGA Survey. This subset of galaxies is in a specific region of the photometric space that low-redshift ($z < 0.2$) galaxies preferentially occupy, defined as the \textit{Primary Targeting Region} in \cite{Mao_2021}. This region is defined by:
\begin{equation}
    (g-r)_o-\sigma_{gr}+0.06(r_o-14)<0.9,\label{eq:colour_cut}
\end{equation}
\begin{equation}
    \mu_{r_o, \mathrm{eff}} + \sigma_\mu -0.7(r_o-14)>18.5,\label{eq:sb_cut}
\end{equation}
where
\begin{equation}
    \mu_{r_o, \mathrm{eff}} = r_o + 2.5\log_{10}(2\pi R^2_{r_o,\mathrm{eff}}).\label{eq:sb_calc}
\end{equation}
In the above equations, $g$ and $r$ are the $g$- and $r$-band magnitudes, and $R_{r_o, \mathrm{eff}}$ is the full width at half maximum radius in the $r$ band. The $1\sigma$ error on a quantity $a$ is quoted as $\sigma_a$, and $\sigma_{gr}=\sqrt{\sigma_g^2+\sigma_r^2}$.
The subscript o, e.g., $r_o$, refers to the extinction-corrected version of the quantity.

There are two main reasons to focus on this subset of spectra in the SAGA Primary Targeting Region. First, the SAGA Survey has obtained high completeness in this region ($\gtrsim 90$\% of galaxies in SAGA's photometric catalogue in this region down to $r_o = 20.7$ have spectra). Second, the galaxy redshift distribution in this region peaks around $z=0.1$ and has few galaxies above $z > 0.25$.  Hence, this allows us to start with a sample that already has low contamination from high-redshift galaxies. 
In fact, using redshifts from the SAGA Primary Targeting Region to train machine learning algorithms has been shown to be effective in selecting low-redshift galaxies \citep{Wu2022, Elise2023}. 
It is worth noting that the SAGA Primary Targeting Region is designed to be colour-complete over a range of environments at $z<0.015$, and at higher redshifts will be biased towards bluer galaxies \citep{Mao_2021,Elise2023}.

Figure~\ref{fig:saga_hexbins} demonstrates the two relevant photometric selections for the SAGA Primary Targeting Region and the corresponding average stellar mass of each bin in the photometric space.  The stellar mass estimates are provided by SAGA and are determined using the $k$-corrected $r$- and $g$-band absolute magnitudes, with a systematic error of $0.2\dex$ \citep{Mao_2021}.
Because this study focuses on low-mass galaxies, we further limit the \textit{spectroscopic calibration sample} to $r_o > 18$ to remove galaxies of $M_*=10^{11}\,\msun$ and above. The blue-shaded region of Figure~\ref{fig:saga_hexbins} highlights our final \textit{spectroscopic calibration sample}, which contains 6,330 galaxies. Of these, 2,318 have stellar masses below $M_*<10^{9}\,\msun$. About 95\% of the redshifts in the \textit{spectroscopic calibration sample}  were first measured by the SAGA Survey team, and the remaining were compiled from the literature (mainly SDSS; see \citealt{Mao_2021} for a complete list).

\subsection{Source and lens samples from the Dark Energy Survey}\label{sec:DESdata}

The Dark Energy Survey \citep[DES;][]{Abbott_2018,  Abbott_2021} is a six-year photometric survey by the DES collaboration at the Cerro Tololo Inter-American Observatory (CTIO), Chile, using the Blanco telescope and the 570-megapixel Dark Energy Camera \citep{Flaugher_2015}.
The survey covers an area of $\sim 5000 \,\mathrm{deg}^2$ of the southern Galactic cap with five broad-band filters, \textit{grizY}.

\subsubsection{Source Sample}
\begin{figure}
    \centering
\includegraphics[width=\columnwidth]{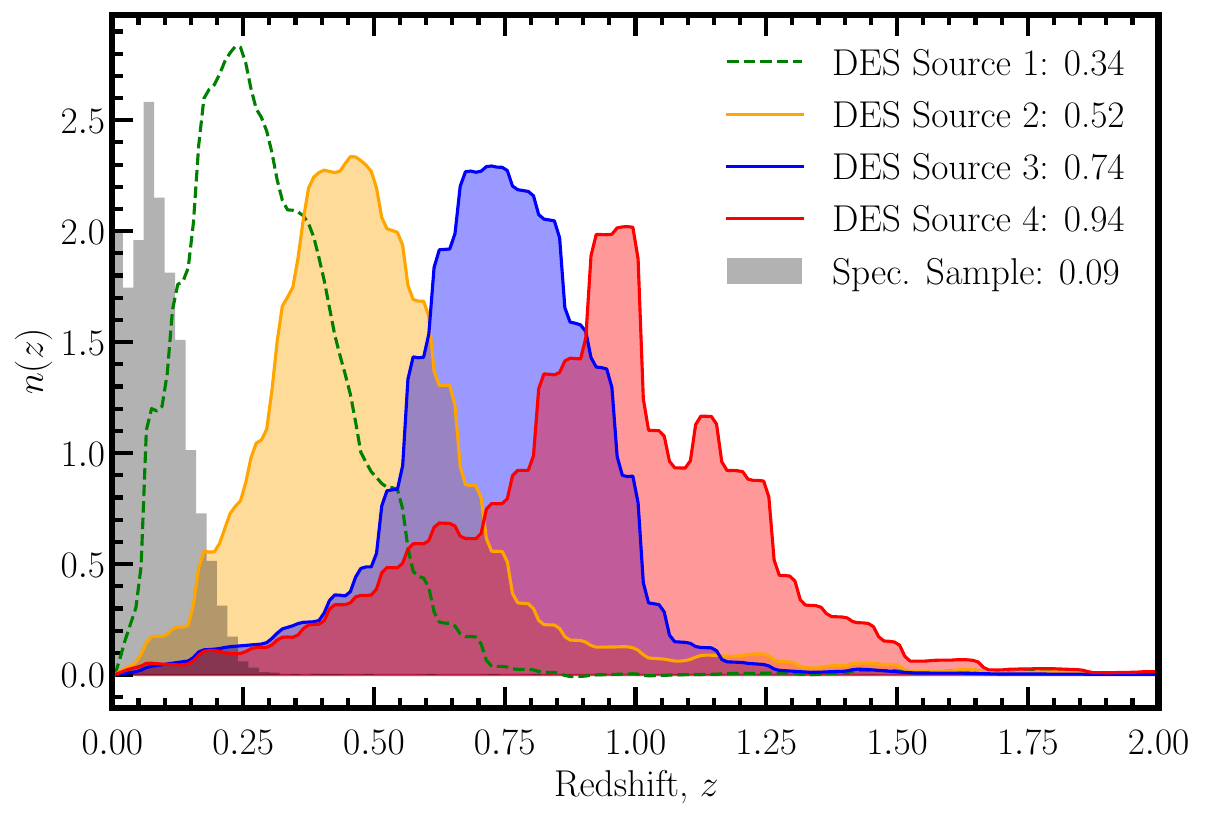}
    \caption{Redshift distributions of the \textit{spectroscopic calibration sample} (grey histogram; Spec. Sample) and the DES Y3 calibrated photometric source samples (coloured), divided into four redshift bins following \citep{Myles_21}.
    The lowest redshift bin of the DES source sample is shown as an unfilled distribution as it is not used in this analysis in order to minimise redshift overlap with the lens sample. The mean redshift, $\bar{z}$, of each sample is quoted in the legend.
    The distribution of the \textit{spectroscopic calibration sample} is normalised to aid visual comparison.}
    \label{fig:nz}
\end{figure}

The background \textit{source sample}, or lensing data, used in this work is taken from the first three years of DES \citep[Y3;][]{Sevilla-Noarbe_2021}, with a footprint of $4143 \,\,\mathrm{deg}^2$.
The catalogue contains 100,204,026 galaxies, with a number density of $5.59 \,\,\mathrm{arcmin}^{-2}$ \citep{Gatti_2021}, and shear measurements determined with the \textsc{metacalibration} pipeline \citep{huff_2017, Sheldon_2017}.
\citet{Myles_21} divided this source sample into four bins and calibrated their probability redshift distributions to per cent level precision, primarily using a SOM method that incorporates overlapping spectroscopic redshifts. The calibrated distributions are displayed in Figure~\ref{fig:nz}.
In this work, we discard the lowest redshift source bins in favour of the three higher DES Y3 source bins in order to minimise overlap with the low-mass foreground sample. The shear bias corrections are calibrated and defined for each redshift bin in \citet{MacCrann2022}.
For the self-calibration of the \textsc{metacalibration} shapes, we use the response factors calculated for each source bin in \citet{Prat_Y3}.
These will both be discussed in more detail in Section~\ref{sec:est}.

\subsubsection{Candidate low-mass sample}\label{sec:des_clms}

The DES second data release \citep[DES DR2;][]{Abbott_2021} includes $\sim 691$ million objects. Since the majority of these objects are not low-mass galaxies, we first construct a \textit{candidate low-mass sample}. 
This initial \textit{candidate low-mass sample} is constructed such that it occupies the same photometric space as the \textit{spectroscopic calibration sample} described in Section~\ref{sec:saga}. As such, we apply the same photometric selection as outlined in \citet[see Section~3.2, 3.6 and 4.1 therein]{Mao_2021} on the DES DR2 catalogue. The properties used below refer to column names from the DES DR2 catalogue, unless otherwise specified. For extinction-corrected magnitudes, we use \texttt{MAG\_AUTO\_*\_DERED}, and for radius, we use \texttt{FLUX\_RADIUS\_*} where \texttt{*} refers to the band in question, e.g., \textit{g, r, i, z}.

\begin{enumerate}

\item We begin by applying DES quality cuts to identify `good' galaxy targets:\label{step:good}
\begin{align*}
    \texttt{IMAFLAGS\_ISO\_R}&=0,\\
    \texttt{FLAGS\_R}&<4
\end{align*}
These are standard \textsc{SExtractor} \citep{SExtractor} flags, mostly removing saturated stars.\\

\item Star--galaxy separation:\label{step:sg_sep} To delineate galaxies and point sources, we use the \texttt{EXTENDED\_COADD} parameter in the $r$-band, defined as \footnote{\label{foot:faq}\url{https://des.ncsa.illinois.edu/releases/dr2/dr2-faq}.}:
\begin{align*}
    &\texttt{EXTENDED\_COADD}=\\
    &(\texttt{SPREAD\_MODEL\_R}+3\times\texttt{SPREADERR\_MODEL\_R} > 0.005)+\\
    &(\texttt{SPREAD\_MODEL\_R}+\texttt{SPREADERR\_MODEL\_R} > 0.003)+ \\
    &(\texttt{SPREAD\_MODEL\_R}-\texttt{SPREADERR\_MODEL\_R} > 0.003),
\end{align*}
and we require galaxies to have \texttt{EXTENDED\_COADD}=3 \citep[e.g.~][]{Sevilla-Noarbe_2021, Abbott_2021}.\\

\item Second star--galaxy separation:\label{step:sg_sep2} For very bright objects ($r_o<$~17), stars can be misclassified as galaxies, and we remove these by excluding objects that jointly satisfy the following criteria \citep{Mao_2021}:
\begin{equation}
    0.7\cdot(r_o +10.2) >
    2.5\log_{10}(2\pi R^2_{r_o,\mathrm{eff}}),\label{eq:missClassify1} 
\end{equation}
\begin{equation}
    g_o-r_o < 0.6,\label{eq:missClassify2}
\end{equation}
\begin{equation}
    r_o < 17.\label{eq:missClassify3}
\end{equation}
  
\item Application of the photometric selection of the \textit{spectroscopic calibration sample} (see Section~\ref{sec:saga} and \citealt{Mao_2021}):\label{step:SAGA} We calculate the surface magnitude for all objects, given by Eq.~\eqref{eq:sb_calc} and apply Eqs.~\eqref{eq:colour_cut} and \eqref{eq:sb_cut} to the DES DR2 catalogue, to extract a low-redshift sample.\\

\item Faint magnitude cut\label{step:mag_cut}: Since the \textit{spectroscopic calibration sample} only extends to a magnitude of $r_o<20.75$, we apply this cut as we do not have spectroscopic information below this magnitude limit.\\

\item Colour bounds\label{step:gr_bounds}: We impose a bounding on $(g-r)_o$ colour to remove objects with poor photometry or that are likely stars, and restrict to the colour range of the \textit{spectroscopic calibration sample}.
These objects are a small proportion of the remaining catalogue, but their anomalously large $\vert(g-r)_o\vert$ colours would affect the SOM training (see Appendix~\ref{sec:SOM_app}).
We retain only objects satisfying:
\begin{equation}
    -1< (g-r)_o < 1.5\,\,.
\end{equation}

\item Bright magnitude cut\label{step:mag_cut2}: Similar to the \textit{spectroscopic calibration sample}, we impose an additional selection beyond those in \citet{Mao_2021} to exclude brighter objects, with larger stellar masses, defined as $r_o > 18$ (Figure~\ref{fig:saga_hexbins}, blue shaded region).

\end{enumerate}
The number of remaining objects present after each section of cuts is shown in Table~\ref{tab:cut-fractions}, which demonstrates less than $\sim 0.2$ per cent of DES DR2 comprises the \textit{candidate low-mass sample}. In Appendix~\ref{sec:repro_app}, we discuss how well-represented the \textit{candidate low-mass sample} is by the \textit{spectroscopic calibration sample}, which has an identical photometric selection.

\begin{table*}
    \centering
    \def\arraystretch{1.1}%
    \caption{Selection criteria applied to the DES DR2 sample to mimic the \textit{spectroscopic calibration sample} selection, and the corresponding percentage of objects retained compared to the full DES DR2 dataset.
    Selection (i) represents DES quality cuts, mainly removing saturated stars based on two standard \textsc{SExtractor} flags.
    Selection (ii) and (iii) introduce star--galaxy separation criteria, based on \citet{Sevilla-Noarbe_2021, Abbott_2021}.
    Step (iv) represents the \textit{spectroscopic calibration sample} selection criteria for a low-redshift sample.
    Step (v) limits the DES data to the available depth of the \textit{spectroscopic calibration sample}, while step (vi) helps remove additional contamination from poor photometry. Finally, step (vii) reduces the stellar mass range by removing the brightest objects.
    From the 691,483,608 objects in the original photometric catalogue, 755,837 remain as the \textit{candidate low-mass sample}. 
    Note that where the cut column displays an inequality, the cut refers to the objects retained.}
    \begin{tabular}{cp{75mm}C{30mm}c}
    \hline
    Index & Description & Selection & \% Retained\\
    \hline
    \ref{step:good} & DES quality cuts & \texttt{IMAFLAGS\_ISO\_R} $=0$ & 99.75\\
     & & \texttt{FLAGS\_R} $<4$ & 99.73\\
    \ref{step:sg_sep} & Identifying likely galaxies & \texttt{EXTENDED\_COADD} $=3$ & 38.56\\
    \ref{step:sg_sep2} & Remove misclassified stars  & eqns.~\eqref{eq:missClassify1}~-~\eqref{eq:missClassify3}  & 38.47\\
    \ref{step:SAGA} & SAGA Primary Targeting Region \citep{Mao_2021} & $(g-r)_o$ cut eq.~\eqref{eq:colour_cut} & 13.28\\
     & & $\mu_{r_o, \mathrm{eff}}$ cut eq.~\eqref{eq:sb_cut}& 7.78\\
    \ref{step:mag_cut} & Limiting to the available depth & $r_o < 20.75$ & 0.17\\
    \hline
    \hline
    \ref{step:gr_bounds} & Colour cut to remove poor photometry / stars & $-1<(g-r)_o<1.5$ & 0.17\\
    \ref{step:mag_cut2} & Manual magnitude cut to reduce stellar mass range & $r_o > 18$ & 0.15\\
     & DES Y3 mask applied & & 0.11\\
    \hline
    \end{tabular}
    \label{tab:cut-fractions}
\end{table*}

\section{Low-mass galaxy selection}\label{sec:som}

In this section, we describe how we characterise the stellar mass and redshift probability distributions of the 
\textit{candidate low-mass sample} using the  \textit{spectroscopic calibration sample}, and further split the photometric data into three separate \textit{low-mass samples}.
Many different methods of nonlinear dimensionality reduction can be employed to impose mappings between our datasets. In this work, we use unsupervised machine learning in the form of a self-organising map \citep[SOM;][]{Kohonen1982, kohonen1990}.
We chose to employ this method based on its success in the photometric redshift calibration, to percent-level accuracy, of state-of-the-art lensing data for cosmological analyses \citep{Masters_2015, Buchs_19, Alacron2020, wright_2020, Myles_21, Hildebrandt_2021, Giannini2022, Sanchez2023}.
This tool allows us to use a limited \textit{spectroscopic calibration sample} to select and characterise subsamples of a large photometric dataset based on some property that is not observed. 
We extract three mass-selected subsamples from the \textit{candidate low-mass sample} and infer their probability stellar mass distribution, $p(M_*)$, as well as their redshift distribution, $p(z)$.

\subsection{SOM training and assignment}\label{sec:SOMtrain}

An overview of the SOM training process is discussed in Appendix~\ref{sec:SOM_app} and for more detail, we refer the reader to \cite{Buchs_19}. 
In brief, we train the SOM using the magnitude, $r_o$, colour, $(g-r)_o$, and surface magnitude, $\mu_{r_o, \mathrm{eff}}$, of a random subset of the \textit{candidate low-mass sample}. Now these 256 cells discretise
and represent the [$r_o$, $(g-r)_o$, $\mu_{r_o, \mathrm{eff}}$]-space of the photometric \textit{candidate low-mass sample}.
Once trained, we populate the SOM with the \textit{candidate low-mass sample}, thereby categorising that data into galaxy types defined by the cells. In order to `label' the cells with redshifts and stellar masses, we then populate with the \textit{spectroscopic calibration sample}. In Appendix~\ref{sec:SOM_app}, we illustrate the results of the trained SOM in terms of the number of photometric and spectroscopic objects, training properties, and inferred properties.

For each cell, $c$, we can now calculate its respective probability distribution, $p(z|c)$, and, $p(M_*|c)$.
For the inferred $M_*$ (and equivalently for the redshift, $z$), we have that:
\begin{equation}
    p(M_* | c) = \frac{1}{A_c}\sum_{\text{obj}\in c}p(M_* | \text{obj}),\label{eq:prob_cell}
\end{equation}
where
\begin{equation}
    A_c = \int_0^\infty \sum_{\text{obj}\in c}p(M_* | \text{obj})\mathrm dM_*,
\end{equation}
and $p(M_* | \text{obj})$ is the probability distribution of the stellar mass for a given calibration object.
Initially, we take $p(M_* | \text{obj})$ to be a delta function centred at the recorded values of the calibration sample.
Once the entire calibration sample is assigned, we convolve a given cell's mass distribution (consisting of delta functions) with a Gaussian such that 
\begin{equation}
    p(M_* | c)\to p(M_*| c)\,\,\ast\,\,\frac{1}{\sigma\sqrt{2\pi}}\exp\left( -\frac{M_*}{2\sigma^2}  \right).\label{eq:prob_conv}
\end{equation}
As our ability to quantify the distribution per cell is limited by spectra, we wish for the spread to reflect additional uncertainty in cells containing few objects.
The variance $\sigma^2$ in Eq.~\eqref{eq:prob_conv} is thus composed of the true uncertainty from \citet{Mao_2021} measurements, $\sigma_\mathrm{true}$, and an additional scatter to account for uncertainty in low spectra count cells, $\sigma_\mathrm{add}$.
We propose a relation
\begin{equation}
    \sigma^2 = \left(1-N_{s,c}^{-1}\right)\sigma_\mathrm{true}^2 + N_{s,c}^{-1}\sigma_\mathrm{add}^2,\label{eq:prob_spread}
\end{equation}
where $N_{s, c}$ is the number of spectra assigned to a cell, $c$.
In this work, we use $\sigma_\mathrm{true} = 0.2\dex$ (see Section~\ref{sec:saga}), and the additional spread is taken to be the $1\sigma$ spread of the full \textit{spectroscopic calibration sample} distribution, $\sigma_\mathrm{add}\sim 0.9\dex$.
In the future, a more rigorous method of accounting for uncertainty in the assignment when dealing with a low number of spectra could be better accounted for by varying the power law and normalisation of Eq.~\eqref{eq:prob_spread}'s dependence on $N_{s, c}$.
With the abundance of new spectra expected in the near future with surveys such as DESI, it is unlikely methods such as this will be required.
For the remainder of this work, we use the procedure outlined in Eqs.~\eqref{eq:prob_conv} and \eqref{eq:prob_spread} for our estimated distributions.

\begin{figure*}
    \centering
    \includegraphics[width=\textwidth]{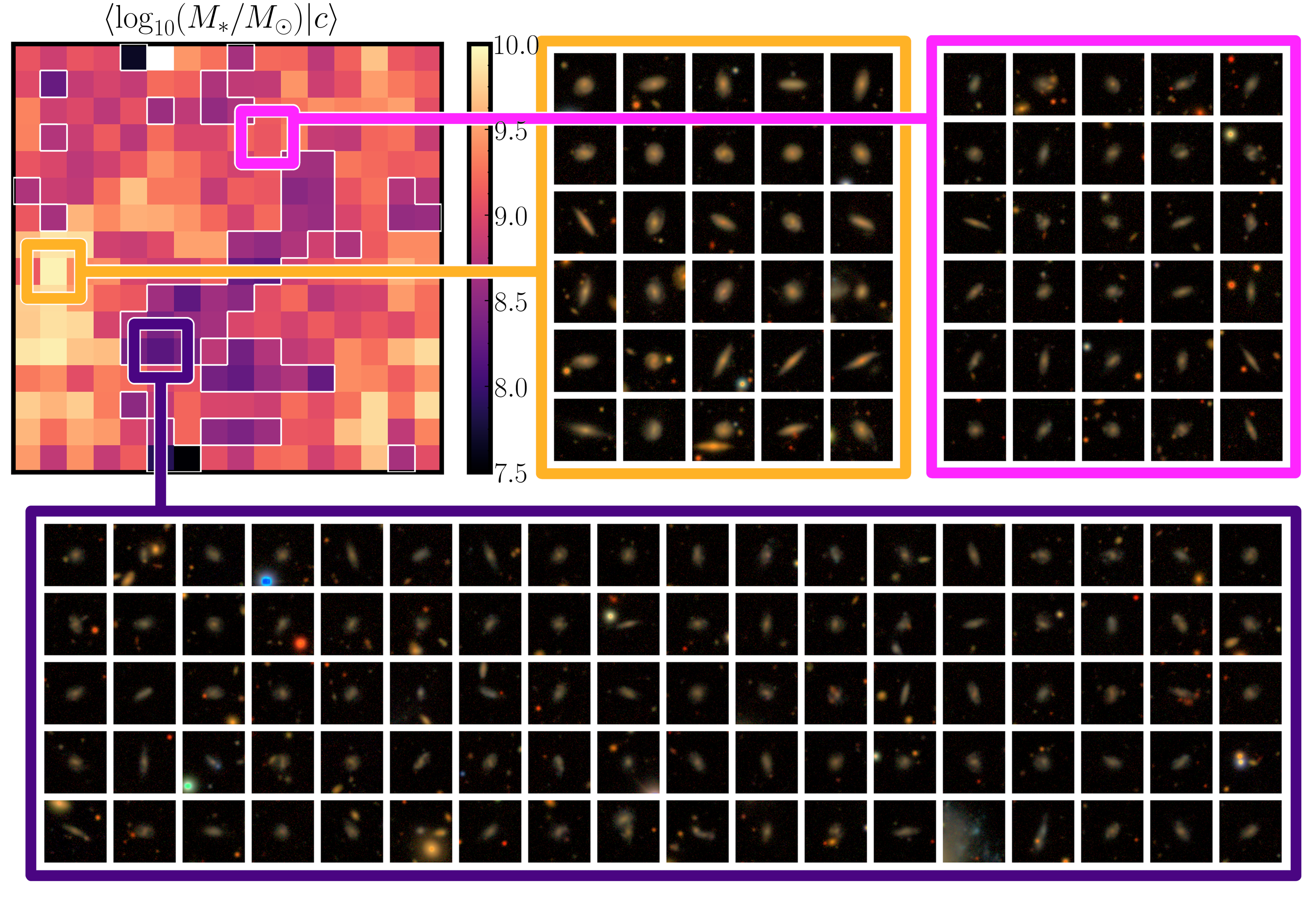}
    \caption{The self-organising map is trained using the \textit{photometric candidate low-mass sample}. Each `cell' in the grid represents a galaxy type (in this case, galaxies that are similar in $r$-magnitude, $(g-r)$ colour and surface brightness). 
    The colour of the cell indicates the average stellar mass of the spectroscopic galaxies in that cell, with corresponding values in the colour bar. To demonstrate the effectiveness of the SOM in sorting galaxies by mass, we choose three cells of different average stellar mass and highlight postage stamp images of random \textit{candidate low-mass sample} galaxies within that cell. Visually, the galaxies in each of these cells have clear morphological differences with $\langle\log_{10}(M_*/\msun)\rangle$ being 9.94 (orange), 9.10 (pink), 8.19 (purple).
    The angular size of each cutout measures $0.4\times 0.4$ arcmin and the galaxy images were produced using the \textsc{DESaccess} cutout service, producing RGB images in the \textit{Lupton} format, hence the red colouring.
    }
    \label{fig:som_cell_test}
\end{figure*}

In Figure~\ref{fig:som_cell_test}, we demonstrate the effectiveness of the SOM in separating and, therefore selecting, a galaxy sample according to its stellar mass. We visually explore the galaxies assigned to different cells using galaxy images produced using the \textsc{DESaccess} cutout service\footnote{\url{https://des.ncsa.illinois.edu/desaccess/cutout}}.
The three selected cells were chosen for a high discrepancy in assigned stellar mass; 90 objects were selected at random from those that fell in the lowest-mass cell, and 30 from the remaining two.
The angular size of every cutout measures $0.4\times 0.4$ arcmin. This demonstrates that the SOM is selecting groups of objects that are visually different, with the lower mass assigned objects ($\langle\log_{10}(M_*/\msun)\rangle = 8.19$)
being smaller and dimmer than the higher mass ($\langle\log_{10}(M_*/\msun)\rangle = 9.94$).

\subsection{Stellar mass bin selection from SOM cells}\label{sec:mass_bins}

We order the SOM cells by mean stellar mass and divide them into three bins, which we chose to define as
\begin{equation}
   \textsc{LOW:}  \,\, \langle\log_{10}(M_*/\msun)\rangle < 8.75\, ,
\end{equation}
\begin{equation}
   \textsc{MID:}  \,\,     8.75\leq\langle\log_{10}(M_*/\msun)\rangle < 9.20\, ,
\end{equation}
\begin{equation}
   \textsc{HIGH:}  \,\,     9.20\leq\langle\log_{10}(M_*/\msun)\rangle\, .
\end{equation}
With the exception of a negligible fraction ($0.1\%$), all galaxies in the \textit{candidate low-mass sample} are assigned to one of these three  \textit{low-mass samples} (LOW, MID, HIGH) based on which SOM cell each galaxy is in.

The probability distributions for a given mass sample, $b$, can be calculated by weighting the distribution for each cell, $c\in b$, by the number of objects assigned to it. For example, in the case of stellar mass, we have
\begin{equation}
    p(M_* | b) = \frac{\sum_{c\in b} N_{c}p(M_*  | c)}{\sum_{c\in b} N_{c}},\label{eq:distrb_true}
\end{equation}
where $N_c$ is the number of objects in a given cell, $c$.
We produce two distinct $p(M_*|b)$ distributions for each mass sample, $b$, by weighting by either the number of spectroscopic calibration objects or the number of candidate low-mass objects in a given cell.
In the case of weighting the low-mass samples, we use the full distributions described by Eqs.~\eqref{eq:prob_cell} through \eqref{eq:prob_spread}; for the calibration weighting we neglect the $N_{s, c}^{-1}$ dependence in Eq.~\eqref{eq:prob_spread}, taking the spread in the convolved Gaussian simply as $\sigma_\mathrm{true}$, as this is the true distribution of spectroscopic calibration objects.

\begin{figure*}
    \centering
    \includegraphics[width=\textwidth]{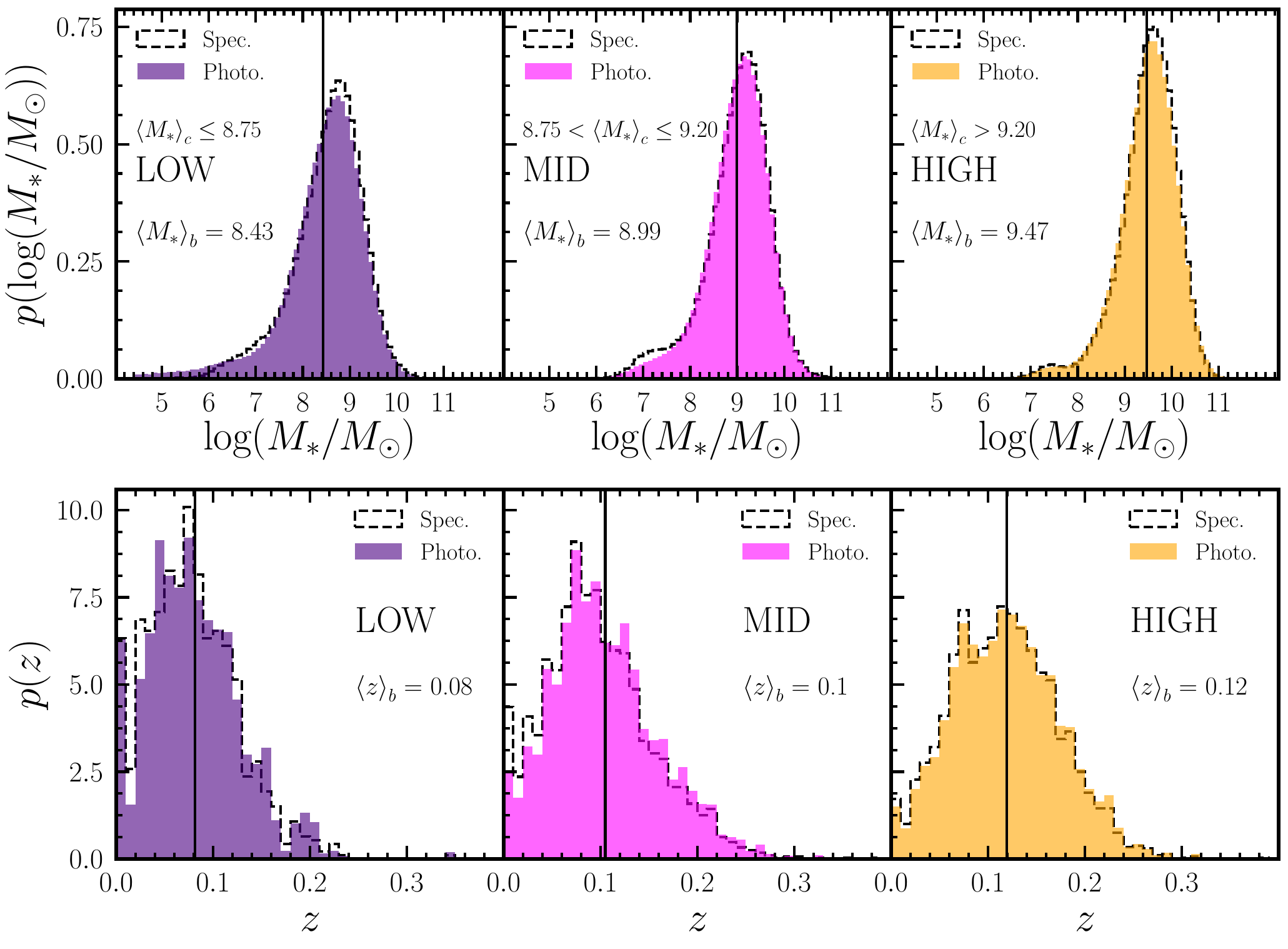}
    \caption{Probability distributions for the stellar mass, $\log_{10}(M_*/\msun)$, (upper panel) and redshift, $z$, (lower panel) for the three \textit{low-mass samples}, each with 146,420 (\textsc{LOW}) 330, 146 (\textsc{MID}), and 275,028 (\textsc{HIGH}) objects.
    The dashed outlines indicate the distributions of the \textit{spectroscopic calibration sample} (here labelled `Spec.') after assignment to a given cell and mass bin. The coloured regions indicate the estimated distributions of the \textit{candidate low-mass sample} (`Photo.'), discussed in Section~\ref{sec:mass_bins}, with the mean indicated as a solid black line. The colour for each sample is used throughout the remainder of this work.}
    \label{fig:true_est_distrb_TOMO}
\end{figure*}

If the \textit{spectroscopic calibration sample} and \textit{low-mass samples} are representative of one another, we would expect the distributions of these two catalogues to be of similar shape.
In Figure~\ref{fig:true_est_distrb_TOMO} we show the final redshift and stellar mass distributions for each of the three mass bins.
The two distributions are closely matched in each of the three bins, suggesting the distribution of the \textit{spectroscopic calibration sample} is representative of the wider population given the same selection criteria.
The mean and median masses of each low-mass sample distribution are displayed in Table~\ref{tab:SH_results}.
We estimate that ${\sim}75$ per cent of the galaxies in the \textsc{LOW} mass sample have $M_* < 10^9\msun$.

\section{Mass profile measurement}\label{sec:ggl}

In this section, we describe the estimator for the excess surface mass density, $\Delta\Sigma(R)$. We present measurements using the \textsc{LOW}, \textsc{MID}, and \textsc{HIGH} \textit{low-mass samples} as the lenses with the DES Y3 background {source sample}. We assume a $\Lambda$CDM cosmology with parameters from \citet{planck18}.

\subsection{Galaxy--galaxy lensing estimator}\label{sec:est}
The presence of a foreground lens induces a shear in the images of background galaxies that, unlike intrinsic ellipticity, does not average to zero when stacked over many sources.
GGL only induces shear tangentially and is computed as the weighted cross-correlation of the shapes of a background \textit{source sample} with the positions of the foreground lens galaxies. The average tangential shear estimator is a summation over all lens--source pairs within a given angular separation, $\theta$, given as
\begin{equation}
    \langle \gamma_{\rm t}(\theta)\rangle = \frac{\sum_\mathrm{ls}\epsilon_{\rm t} w_\mathrm{ls}}{\sum_\mathrm{ls}w_\mathrm{ls}},\label{eq:gamma_t}
\end{equation}
where $\epsilon_{\rm t}$ is the tangential component of the ellipticity of each source galaxy and $w_\mathrm{ls} = w_\mathrm{l}w_\mathrm{s}$ is the combined weight of each lens and source pair, respectively.

Our physical quantity of interest is excess surface density, or the differential profile of the projected mass density of a halo, defined as 
\begin{equation}
    \Delta\Sigma=\Sigma(R)-\overline{\Sigma}(<R)\,,\label{eq:esd}
\end{equation}
where $\Sigma(R)$ is the projected surface density at a radial distance, $R$, and the average surface mass density within a radius, $R$, is
\begin{equation}
    \overline{\Sigma}(<R)=\frac{2}{R^2}\int_0^R\Sigma(R')R'\mathrm{d}R'\,.
\end{equation}
The excess surface density can then be related to the average tangential shear, $\langle \gamma_\mathrm{t}(\theta)\rangle$, by
\begin{equation}
    \Delta\Sigma=\frac{\langle \gamma_\mathrm{t}(\theta)\rangle}{\overline{\Sigma_{\mathrm{crit}}^{-1}}} \, ,\label{eq:gammaTratio}
\end{equation}
at a projected separation $\theta= R/\chi(z_{\rm l})$, where $\chi(z_{\rm l})$ is the comoving distance to the lens and $\Sigma_{\mathrm{crit}}^{-1}$ is the critical surface mass density. For a source--lens pair at angular diameter  distances $D_{\rm s}$, $D_{\rm l}$, respectively and a relative distance of $D_{\rm ls}$, it is defined as 
\begin{equation}
    \Sigma_{\mathrm{crit}}^{-1}=\frac{4\pi G}{c^2}\frac{D_{\rm l} D_{\rm ls}}{D_{\rm s}} \, .
\label{eq:singleSigCrit}
\end{equation}
For a source sample with a redshift distribution $n(z_{\rm s})$, and a lens sample with $n(z_{\rm l})$, the average comoving\footnote{Note that Eq.~\eqref{eq:singleSigCrit} is the \textit{proper} $\Sigma_\mathrm{crit}^{-1}$, and conversion to the comoving quantity requires an additional prefactor of $(1+z_\mathrm{l})^2$ \citep[see Appendix C1 of][]{Dvornik18}.} inverse critical surface density can be calculated as
\begin{equation}
    \overline{\Sigma_{\mathrm{crit}}^{-1}}=\frac{4\pi G}{c^2}\int_0^\infty n(z_{\rm l})(1+z_{\rm l})^2D(z_{\rm l})\mathrm{d}z_{\rm l}\int_{z_{\rm l}}^\infty n(z_{\rm s})\frac{D(z_{\rm l}, z_{\rm s})}{D(z_{\rm s})}{\mathrm d}z_{\rm s}.\label{eq:sigCrit}
\end{equation}
Note that redshift distributions are normalised such that $\int_0^\infty n(z)\mathrm{d}z=1$. 

Weak lensing shape measurements are imperfect, but two corrections can account for biases incurred.
The DES Y3 \textsc{metacalibration} shape method self-calibrates the average shear measurements to avoid noise and selection biases via a response factor \citep{Sheldon_2017, Gatti_2021}.
The total ensemble response, $\overline{R}$, is the sum of the shear, $\overline{R_{\gamma}}$, and selection, $\overline{R_{\rm s}}$, responses. 
We use the values for $\overline{R}$ computed for each of the DES Y3 source redshift bins in \citet{Prat_Y3} and \citet{Amon_Y3}.
The average shear of an ensemble is corrected for using image simulations.
The average source multiplicative bias correction, $\overline{m}$, is defined for each DES Y3 redshift bin in \citet{MacCrann2022}.
Therefore, an estimator for the comoving excess surface density for a source redshift bin over a sample of lens galaxies is\footnote{
For source bins 2, 3, and 4 respectively we take $\overline{R} = 0.7266, 0.7014, 0.6299$, and $\overline{m} = -0.020, -0.024, -0.037.$
}
\begin{equation}
\Delta \Sigma_{\rm l}(R) = \frac{1}{\overline{R}(1+\overline{m})} \langle \gamma_\mathrm{t}(\theta)\rangle(\overline{\Sigma_{\rm crit}^{-1}})^{-1} \, .
\label{eqn:ds}
\end{equation}
Note that in computing the average tangential shear, the source galaxies are weighted by an inverse-variance weight, $w_{\rm s}$, described in \citet{Gatti_2021} and we assume the lens galaxies are of equal weight, $w_{\rm l}=1$.

In order to remove any additive systematics in the lensing sample, we subtract a measurement of the excess surface density around random lens positions within the DES Y3 mask, $\Delta \Sigma_{\rm r}$.
This is predicted to be consistent with zero in the absence of an additive bias \citep{Mandelbaum2005, Mandelbaum2013, Singh2017}.
We apply a second correction to account for the overlap between the selection of photometric source and lens distributions, which would result in a biased estimate of the excess surface density. 
Given that the lens sample is at low redshift and is well separated along the line of sight, we expect this correction to be small.
This boost factor correction, $B(R)$, is computed using random, un-clustered lens positions with the same selection as the true lens sample \citep[e.g.,][]{Sheldon2004, Amon_2018}.
The boost correction, $B(R)$ is given by 
\begin{equation}
B(R) = \frac{\sum_{\rm ls} w_{\rm l} w_{\rm s}}{\sum_{\rm rs} w_{\rm r} w_{\rm s}} \, ,
\label{eqn:boost}
\end{equation}
where $w_{\rm r}$ corresponds to the random lens galaxy weights, which are assumed to be unity, $w_{\rm r}=w_{\rm l}=1$. 
Finally, the overall estimator is
\begin{equation}
  \Delta \Sigma(R) = B(R) \Delta \Sigma_{\rm l} (R) - \Delta \Sigma_{\rm r} (R) \, .
\label{eqn:dsbr}
\end{equation}
For additional details on lensing validation and corrections, see Appendix~\ref{sec:lens_app}.

\subsection{Galaxy--galaxy lensing measurements}\label{sec:ggl_measurement}
The three \textit{low-mass samples} defined in Section~\ref{sec:mass_bins} act as the lens galaxies; the source sample comprises the three higher-redshift bins from DES Y3 (see Section~\ref{sec:DESdata}). For each lens--source bin pair, we compute the excess surface mass density profiles following the estimator defined in Section~\ref{sec:est}.

We calculate the average tangential shear measurements, $\langle \gamma_\mathrm{t}(\theta)\rangle$, using the software package \textsc{TreeCorr}\footnote{\url{https://rmjarvis.github.io/TreeCorr}} \citep{Jarvis_2004}\footnote{We use the count-shear correlations class, \textsc{NGCorrelation}, and set \textsc{TreeCorr}'s `binslop' parameter to zero in our measurements.} for each lens--source bin pair. To estimate the uncertainty, we compute a jackknife covariance by dividing the lens
galaxies and random points into 150 regions using the \textsc{kmeans3}
algorithm. Given the $\sim 4150$deg$^2$ footprint area, this yields regions of approximately $5\,\,\mathrm{deg}^2$ or $\sim 300\,\mathrm{arcmin}$ of length assuming a square geometry, compared to $100\,\mathrm{arcmin}$, the largest angular scale measured. We validate that the errors are stable to the number of jackknife patches.

The average tangential shear measurements are converted to stacked excess surface mass density profiles  following Eq.~\eqref{eqn:ds}, using the \textit{low-mass sample} redshift distributions, $n(z_{\rm l})$ estimated in Section~\ref{sec:mass_bins} and the DES Y3 source redshift distributions,  $n(z_{\rm s})$.
As the uncertainty on the source $n(z_{\rm s})$ are calibrated to be per cent level \citep{Myles_21}, we neglect their propagation into the final uncertainty  on $\Delta\Sigma(R)$.
For this analysis, we verify that the measurements are relatively insensitive to variations in the estimated lens distributions $n(z_{\rm l})$ given the reduced lensing efficiency,  $D_{\rm ls}/D_{\rm s}$, of a low-redshift lens sample and the current statistical power of the data (for example, an error as large as $\sim20$ per cent on the mean $z_{\rm l}$ causes a few per cent bias to the lensing efficiency). 

For each of the source--lens bin pairs, we compute an average random signal, $\Delta \Sigma_{\rm r} (R)$ and an average boost signal, $B(R)$, and correct the excess surface mass density following Eq.~\eqref{eqn:dsbr}.
We describe these corrections in more detail in Appendix~\ref{sec:lens_app}.
The random signal is consistent with zero, and the boost signal is at most a two per cent correction. We ignore the uncertainty on each of these corrections in our final measurement.

\begin{figure*}
    \centering
    \includegraphics[width=\textwidth]{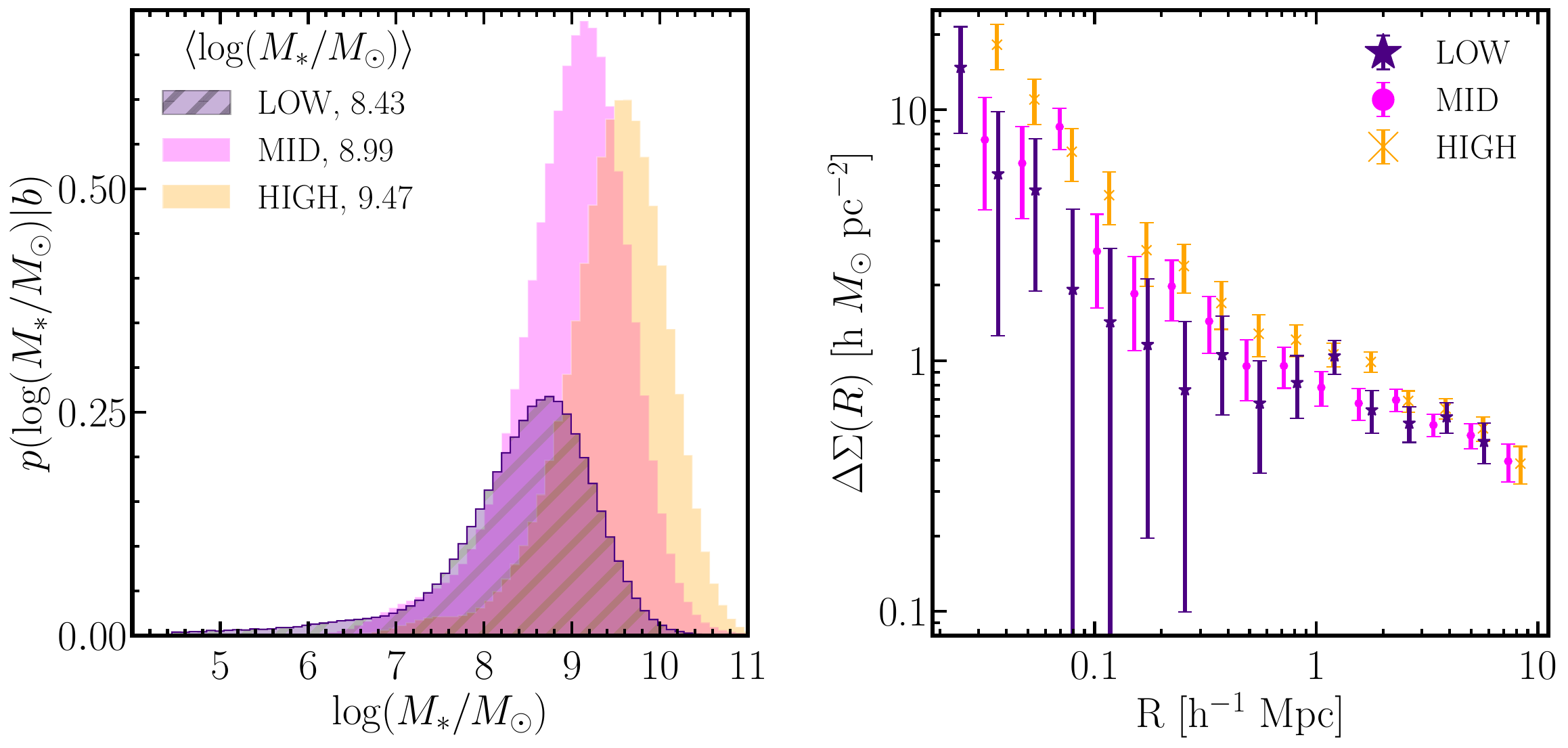}
    \caption{
    Left: Stellar mass probability distributions for the \textsc{LOW} (purple), \textsc{MID} (pink), and \textsc{HIGH} (orange) \textit{low mass samples}, produced by selecting SOM cells for which $\langle\log_{10}(M_*/\msun)\rangle<8.75$, $8.75\leq\langle\log_{10}(M_*/\msun)\rangle<9.20$, and $\langle\log_{10}(M_*/\msun)\rangle>9.20$, respectively. The corresponding mean stellar masses for each distribution are $\langle \log_{10}(M_*/\msun)\rangle=$8.43, 8.99, and 9.47.
    For visibility, the distributions for \textsc{LOW}, \textsc{MID}, and \textsc{HIGH} have been scaled by the number of photometric objects with $146420, \,\,330146$, and $275028$ objects.
    Right: The final calibrated and corrected excess surface mass density measurements, $\Delta\Sigma(R)$ for each sample. As a measure of the signal-to-noise ratio, we calculate a $\chi^2$ comparison with a null signal, producing S/N of 13.6, 22.9, and 27.6 for the \textsc{LOW, MID} and \textsc{HIGH} bins respectively.}
    \label{fig:dual_TOMO_and_BLC}
\end{figure*}

For each lens bin, the measurements are computed using DES Y3 source bins 2, 3, and 4, and are combined using an inverse-variance average, following \citet{Amon_2022}. The right-hand panel of Figure~\ref{fig:dual_TOMO_and_BLC} displays the final stacked $\Delta\Sigma(R)$ signals for the three \textit{low-mass samples}, with their corresponding stellar mass distributions shown in the left-hand panel. The three samples are at roughly the same redshift, so the increasing amplitudes of the lensing signals from the \textsc{LOW} to \textsc{HIGH} bin already indicate an increasing average halo mass.

We compute the signal--to--noise (S/N) ratio of the measurement from each sample based on a $\chi^2$ comparison with a null signal
\begin{equation}
    \mathrm{S/N} = \sqrt{\mathbf{D}\,\, \mathbf{C^{-1}}\,\, \mathbf{D}},
\end{equation}
where $\mathbf{D}$ is our measurement. An unbiased estimate of the inverse covariance matrix, $\mathbf{C^{-1}}$, is calculated by applying a Hartlap correction \citep{Hartlap2007} to the jackknife covariance $\mathbf{C_*}$ as
\begin{equation}
    \mathbf{C^{-1}} = \frac{n - p - 2}{n - 1}\mathbf{C^{-1}_*},
\end{equation}
where $n$ is the number of jackknife patches, and $p$ is the number of elements in our data vector.
We find a S/N$>13$ for the lensing measurement of the \textsc{LOW} mass sample, with the \textsc{MID} and \textsc{HIGH} mass samples' measurements of S/N$>22$. The values are quoted in Table~\ref{tab:SH_results}.  
\begin{table*}
    \centering
    \caption{
    Summary table of the properties of each \textit{low-mass sample}, \textsc{LOW}, \textsc{MID}, and \textsc{HIGH}, and the corresponding model constraints. From left to right, the columns represent the sample name, the total number count of galaxies in the sample, the estimated stellar mass distribution of the sample, the signal-to-noise ratio of its lensing measurement, and the mean and median posterior on the halo mass and satellite fraction obtained from fits to the measurements.
    The quoted errors on the mean stellar masses represent one standard deviation of the distribution, whilst the quoted error on the median stellar masses corresponds to the 84th and 16th percentiles. The signal-to-noise is calculated through a $\chi^2$ comparison with a null model.
    }
    \def\arraystretch{1.5}%
    \begin{tabular}{c|c|ccc|ccccc}
        \hline
        Sample & Count & \multicolumn{3}{c}{Measurement} & \multicolumn{5}{c}{Simulation-based model} \\
        \hline
        & & \multicolumn{2}{c}{Stellar Mass, $\log_{10}(M_*/\msun)$} & S/N & \multicolumn{2}{c}{Halo Mass, $\log_{10}(M_h/\msun)$} & \multicolumn{2}{c}{Satellite fraction}  & $\chi^2_\nu$\\
        && Mean & Median & & Mean & Median & Mean & Median \\
        \hline
        \textsc{LOW} &146420& 8.43$\pm$0.88 & $8.52\substack{+0.57 \\ -0.76}$ & 13.6 &$10.60\pm0.29$& $10.67\substack{+0.23 \\ -0.39}$ &$0.223\pm0.045$& $0.219\substack{+0.039 \\ -0.036}$ & 0.97\\
        \textsc{MID} &330146& 8.99$\pm$0.68 & $9.02\substack{+0.50 \\ -0.64}$ & 22.9  &$10.96\pm0.20$& $11.01\substack{+0.14\\ -0.27}$& $0.176\pm0.021$ & $0.175\substack{+0.022 \\ -0.019}$ & 2.09 \\
        \textsc{HIGH} &275028& 9.47$\pm$0.64 & $9.49\substack{+0.50 \\ -0.58}$ & 27.6  &$11.35\pm0.14$& $11.40\substack{+0.08 \\ -0.15}$ & $0.219\pm0.022$ &  $0.217\substack{+0.021 \\ -0.020}$ & 0.95\\
        \hline
    \end{tabular}
    \label{tab:SH_results}
\end{table*}

\section{Modelling dark matter profiles}\label{sec:Modelling}

In Section~\ref{sec:model}, we describe the simulation-based approach that we adopt to model the lensing mass profile measurements. Section~\ref{sec:mass} presents the resulting model fits to the measurement for each \textit{low-mass sample} and infers the corresponding halo mass distributions. In Appendix~\ref{sect:NFW} we discuss an alternative analytic approach using a simple NFW to model the inner regions of the measurement.

\subsection{Simulation-based model approach}\label{sec:model}

To begin, we propose a simple power-law relation, with a log-normal scatter, between the stellar mass of an object and its surrounding halo mass, $\log_{10}M_{\rm halo}$, given by $\log_{10}M_* = \alpha \log_{10}M_{\rm halo}+\beta+N(0, \sigma) $. This is appropriate for the low-mass end ($\log_{10}M_{\rm halo}<10^{12}$) of the SHMR, which is the mass range that our data is most sensitive to. $N(0,\sigma)$ describes a normal distribution with a mean of zero and a standard deviation of $\sigma$. We parameterise the relation in terms of the slope, $\alpha$, and a typical log stellar mass, $\beta_{11}$ for galaxies that populate a halo mass of $10^{11} \msun$ as
\begin{equation}
    \log_{10}M_*=\alpha\left(\log_{10}M_{\rm halo} - 11 \right)+\beta_{11}+N(0,\sigma)\, .\label{eq:shmr_model}
\end{equation}

For this analysis, we use particle data from the CDM N-body simulation, the Small MultiDark Planck simulation \citep[\textsc{SMDPL,}][]{Klypin:2014kpa}, which assumes a \textit{Planck} cosmology. \textsc{SMDPL} is a $400$ Mpc $h^{-1}$ box of $3840^3$ particles, with a particle mass resolution of $9.63\times 10^{7} \msun$ $h^{-1}$. We use the redshift $z=0$ snapshot in the simulation, primarily because it is publicly available\footnote{\url{https://www.cosmosim.org/}}. Given the redshift range of our stellar mass sample, we expect minimal deviations from $z=0$ halo properties.
The high resolution of the simulation allows us to reliably construct the $\Delta\Sigma(R)$ profiles around haloes with masses $M_{\rm halo}>10^{10.5}\msun$ $h^{-1}$.
For a given SHMR we populate each halo in the simulation with a stellar mass based on our model (Eq.~\eqref{eq:shmr_model}) and then sample over the stellar mass distributions presented in Figure~\ref{fig:true_est_distrb_TOMO} to forward model the expected lensing signal, $\Delta \Sigma$, for a given set of model parameters.
Note that we assign stellar masses using the peak halo mass, $M_\mathrm{peak}$, which is the peak virial mass that a dark matter halo obtains over its entire history. We chose this quantity as it is known to be a more accurate tracer of the stellar mass compared to the current mass of the halo, which can be misestimated due to the tidal stripping of the outer regions of the galaxy \citep{Reddick:2012qy}. We use halo catalogues provided for SMDPL that are generated using the \textsc{rockstar} halo finder \citep{Rockstar} and the consistent trees algorithm \citep{Behroozi2013} which is used to extract the histories of haloes in the simulation to build merger trees.

Naturally, some fraction of our galaxy sample will be satellites and will therefore reside in subhaloes within the virial radius of some larger halo. These objects lose bounded material from their outskirts due to tidal effects once they fall into the host. As a result, their one-halo term is typically truncated, unlike an isolated object.
In the simulation, we know \textit{a priori} which objects are satellites, therefore we can vary the fraction of these that are included in our sampling of the stellar mass distribution, and observe this effect on the $\Delta \Sigma$ profiles. The matter distribution of the larger haloes within which these subhaloes reside will leave an imprint on the excess surface density as measured around the satellite, along with the satellite's own tidal truncation. Note that this fraction of satellites, $f_\mathrm{sat}$, may differ from that predicted by an unbiased sampling of a pure CDM universe \citep[e.g.,][]{Wechsler_2018, Voice2022}, so we include this quantity as an additional parameter in the analysis.

We measure the projected mass density, $\Sigma(R)$, by counting the halo--particle pairs in 30 logarithmic radial bins between $0.01 < R < 10$ Mpc $h^{-1}$ over a projected length of $1$ Mpc $h^{-1}$. The excess surface density is then computed following Eq.~\eqref{eq:esd}.
To assign central galaxies, we require that the parent ID is \textsc{pid}$ = -1$, following \textsc{rockstar}\footnote{\url{https://bitbucket.org/gfcstanford/rockstar}}. Second, we limit central haloes to those that satisfy the condition that the ratio of their peak mass to current mass, $M_\mathrm{peak}/M_\mathrm{today} < 1.01$, in order to account for `splashback' (or backsplash) haloes that will also lose mass from within their own virial regions due to tides within the host. Splashback haloes are objects that are close to their first apocenter in their orbits \citep{Diemer2014, Adhikari14}, but may appear outside the virial radius as the splashback radius/surface, which delineates the bounded region around a halo, may be larger \citep{More2015}. These haloes are in fact orbiting subhaloes but are incorrectly tagged as centrals in a virial-radius-based definition\footnote{Note that we use a simple method based on the ratio of the peak mass to current mass ratio to isolate splashback objects, however a more detailed analysis based on subhalo or particle orbits similar to \cite{Mansfield:2017,2017ApJS..231....5D} may be used.}. A detailed discussion on how this assignment affects clustering properties can be found in \cite{Mansfield:2017}. After isolating the centrals, the remaining haloes are tagged as subhaloes. Having separated haloes into centrals and satellites we can create samples with varying fractions of each and measure their stacked lensing profiles. 

Given the stellar mass distributions of our samples extend below $\log (M_*/\msun)=8.0$, it is important to consider halo masses lower than the resolution limit of the simulation. To estimate the $\Delta \Sigma$ profile for these haloes, we use the simulation-stacked measurements of $\Delta \Sigma$ as a function of mass in ten mass bins centred at between $10.5 < \log [M_{\rm halo}/(\msun/h)] < 15 $ and ten $f_{\rm sat}$ bins between $0 < f_{\rm sat} < 0.8$ to extrapolate down to $M_{\rm halo}=10^9 \msun/h$.

Using the simulation populated with the estimated stellar mass distribution, we fit the excess surface mass density measurements with the stacked simulated profiles, varying the three SHMR model parameters defined in Eq.~\eqref{eq:shmr_model} as well as the satellite fraction. For each of these parameters, we use uniform and uninformative priors defined in Appendix~\ref{sec:SHMR_app}. We sample the parameter space using an MCMC sampler, \textsc{Emcee} \citep{Emcee}. The posteriors are summarised in Appendix~\ref{sec:SHMR_app}.

\subsection{Halo mass estimates}\label{sec:mass}

\begin{figure*}
    \centering
    \includegraphics[width=\textwidth]{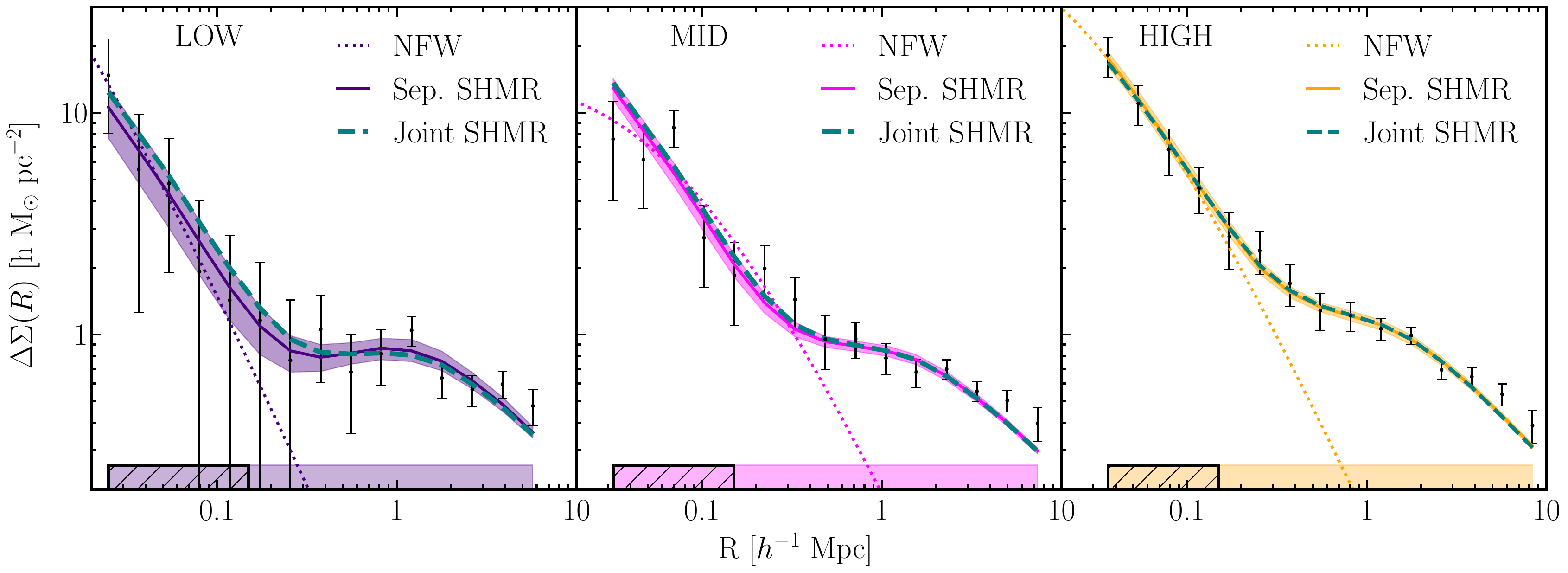}
    \caption{Model fits to the excess surface mass density measurements, $\Delta\Sigma(R)$, in each of the \textsc{LOW}, \textsc{MID} and \textsc{HIGH} mass samples, from left to right panels.
    The median simulation-based profile is indicated as the solid line in purple, pink, and orange respectively, with the $16^\mathrm{th}$ to $84^\mathrm{th}$ percentile bounds shaded around it. 
    The dotted lines indicate the best-fitting NFW model to the inner parts of the profiles.
    The filled bars along the bottom of each panel indicate the region of data constrained by the simulation-based model (coloured fill), and the NFW model (hatched fill).
    This comparison is discussed in Appendix~\ref{sect:NFW}.
    The teal dashed lines indicate the median profiles from the jointly constrained SHMR using all three mass bins.
    }
    \label{fig:final_fits}
\end{figure*}

\begin{figure*}
    \centering
    \includegraphics[width=\textwidth]{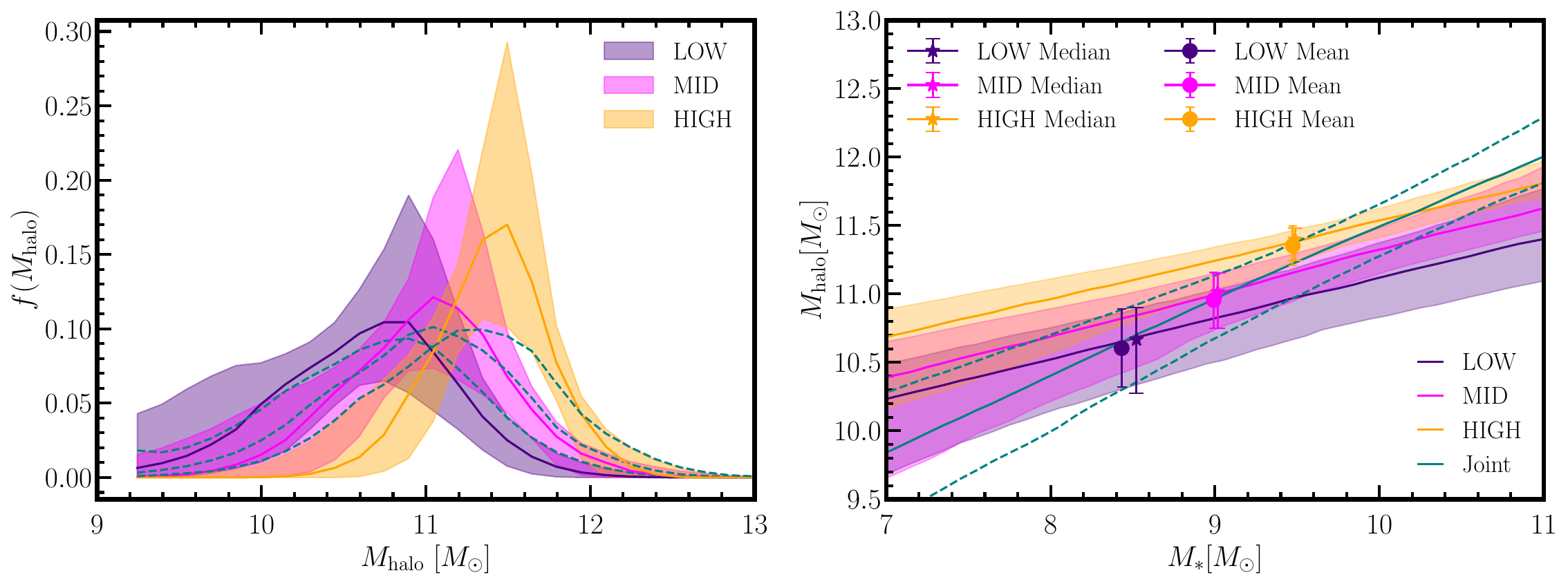}
    \caption{Left: The halo mass distributions, $f(M_{\rm halo})$ derived from the simulation-based model fits to the measurements of the \textsc{LOW} (purple), \textsc{MID} (pink), and \textsc{HIGH} (orange) mass samples.
    The bands show the $16^{\rm th}$ and $84^{\rm th}$ percentile of the halo mass distribution, obtained by populating haloes in SMDPL with stellar masses using the posteriors of the SHMR parameters and drawing observed stellar mass distributions. 
    The dashed teal lines correspond to the distributions drawn from the joint fits, with peak orders following \textsc{LOW}, \textsc{MID}, and \textsc{HIGH}.
    Right: Constraints on the relation between the mean halo mass, $M_{\rm halo}$, for a given stellar mass, $M_{*}$. The bands show the 16$^{\rm th}$ and 84$^{\rm th}$ percentile of the mean relation obtained by sampling the posteriors of the SHMR parameters. Note that the mean halo mass given a stellar mass is given by the SHMR convolved with the halo mass function, as explained in Section~\ref{sec:Modelling}. The shaded bands correspond to individual fits to the three stellar mass bins, and the region interior to the dashed lines correspond to joint fits for the three bins. The solid lines correspond to the median of the posteriors in each case.
    }
    \label{fig:msmh}
\end{figure*}

We fit the lensing measurements, $\Delta \Sigma(R)$, of each of the \textsc{LOW}, \textsc{MID}, and \textsc{HIGH} mass bins separately by varying the parameters of the SHMR and the satellite fraction. The median  $\Delta \Sigma(R)$ profile obtained from each fit is shown in Figure~\ref{fig:final_fits} for each mass bin as a solid line in purple, pink, and orange, respectively. The shaded bands indicate the $1\sigma$ profiles obtained by sampling from the posterior distribution of the parameters. We find good fits to the data, with reduced chi-squared, $\chi^2_\nu$, quoted for the median fits in Table~\ref{tab:SH_results}. 

To best constrain the model parameters, we also perform a joint fit to all three mass bins simultaneously. We use the same SHMR parameters for the three bins, but let the satellite fractions vary independently. The posteriors for the individual and joint fits are shown in Figure \ref{fig:shmr_posteriors}. This provides consistent results to the fit using each mass bin separately, as is indicated by the teal dashed lines in Figure~\ref{fig:final_fits}. In addition, we fit an analytic NFW profile to the inner regions of the measurements (hatched region in Figure~\ref{fig:final_fits}). The best fit using this approach is illustrated as the dotted line, with the comparison discussed in Appendix~\ref{sect:NFW}.

The analysis of each mass sample gives a posterior of the underlying halo mass distribution. The median of each posterior is shown in the left-hand side of Figure~\ref{fig:msmh} for the \textsc{LOW}, \textsc{MID} and \textsc{HIGH} mass samples, and the 1$\sigma$ constraint from the posterior is indicated by the shaded region. These distributions correspond to the distribution of $M_{\rm peak}$ that fit the lensing measurements, obtained by sampling the model parameters. The mean and medians of the halo mass distributions are quoted in Table~\ref{tab:SH_results}, along with the mean and median constraints on the satellite fractions. The median of the posterior mass distributions for the joint fits are also shown in the left panel as teal dashed lines.

On the right-hand side of Figure~\ref{fig:msmh} we show the inferred median (mean) halo mass against the median (mean) stellar mass as stars (circles) for each mass bin. The coloured bands denote the 1$\sigma$ region of the expectation value of the mean halo mass given a fixed stellar mass, $\left<\log M_{\rm halo}|\log M_\ast\right>$, in a $\Lambda$CDM universe, derived from the SHMR posteriors. In particular,
\begin{equation}
\left<\log M_{\rm halo}|\log M_\ast\right>= \int \log M_{\rm halo} P(\log M_{\rm halo}|\log M_\ast)d\log M_{\rm halo}\,
\end{equation}
where $P(\log M_{\rm halo}|\log M_\ast)$ is the probability of getting a halo mass $M_{\rm halo}$ at stellar mass $M_\ast$, it is given by, 
\begin{equation}
P(\log M_{\rm halo}|\log M_\ast)= 
\frac{P(\log M_\ast|\log M_{\rm halo})\phi(M_{\rm halo})}{\int P(\log M_\ast|\log M_{\rm halo})\phi(M_{\rm halo})d \log M_{\rm halo}}\,,
\end{equation}
and $P(\log M_\ast| \log M_{\rm halo})$ follows from the relation  
\begin{equation}
P(M_\ast|M_{\rm halo})=\frac{1}{\sqrt{2\pi\sigma^2}}\rm{exp}(-\frac{(M_\ast-\mu)^2}{2\sigma^2}) \,,
\end{equation}
with $\mu$ and $\sigma$ derived from the mean and scatter of the SHMR in Eq.~\eqref{eq:shmr_model}. $\phi(M_{\rm halo})=dN/d\log M_{\rm halo}$ is the halo mass function; while this can be directly inferred from the simulation, to create the right-hand panel we used the parameterisation of \cite{Sheth:1999mn} to extrapolate the relation consistently to low masses.  
The constraint from the joint fit is also shown. 
The joint fit predicts a slightly lower mean mass for the \textsc{HIGH} sample but is statistically consistent with the individual fits. 

\begin{figure*}
    \begin{minipage}{\columnwidth}
        \centering
        \includegraphics[width=\columnwidth]{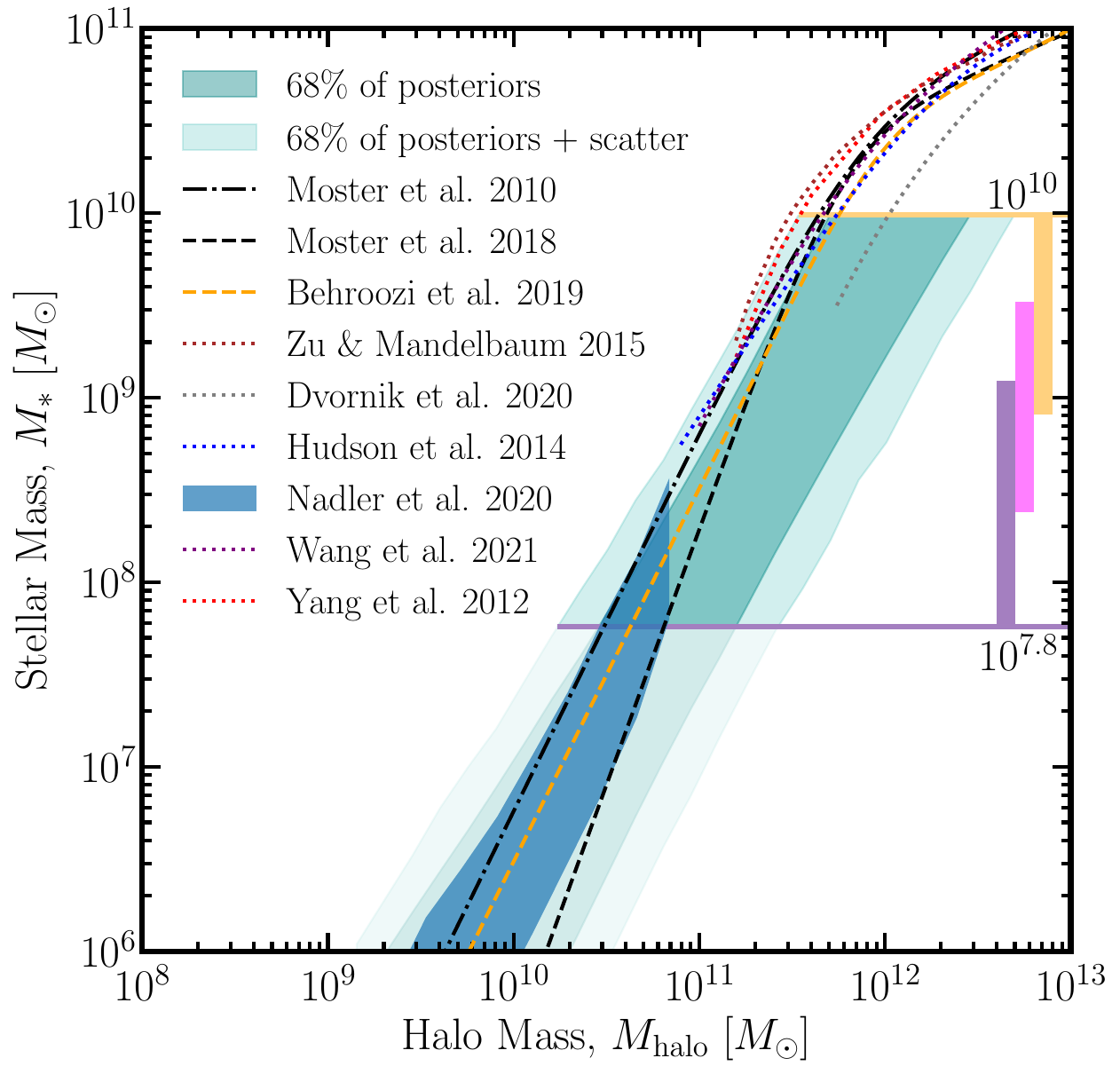}
    \end{minipage}%
    \begin{minipage}{\columnwidth}
        \centering
        \includegraphics[width=0.96\columnwidth]{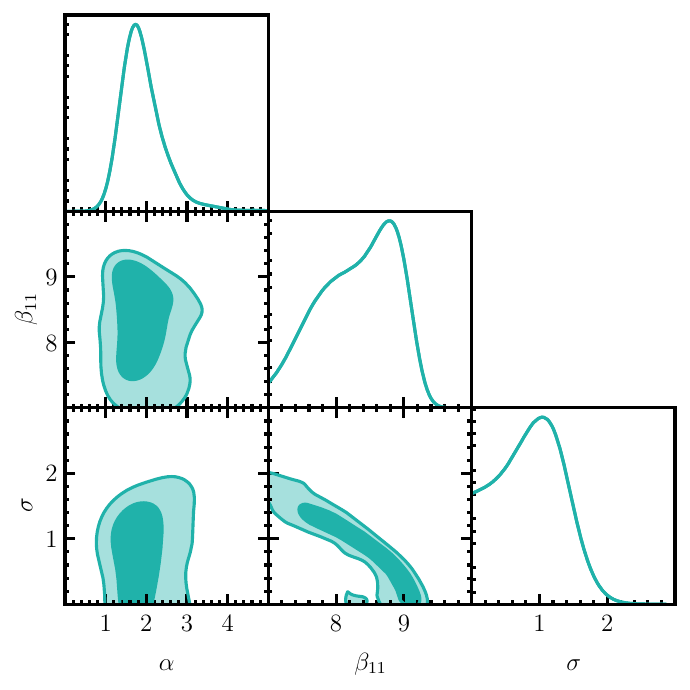}
    \end{minipage}
    \caption{Left: The SHMR, $M_*$ vs. $M_{\rm halo}$, produced from the joint constraint over all three mass bins using Eq.~\eqref{eq:shmr_model} in teal.
    The darker teal-shaded region corresponds to the 68\% region of possible parameters for the SHMR without including the intrinsic scatter, $\sigma$ while the lighter region includes that term.  The horizontal lines indicate the 16th percentile bound on the LOW sample to the 84th percentile bound on the HIGH sample, indicating the main stellar mass range ($10^{7.8}-10^{10}$) probed in this work. The shaded vertical bars indicate the 68\% region of the \textsc{LOW} (purple), \textsc{MID} (pink), and \textsc{HIGH} (orange) mass samples.
    Overlaid are constraints from existing literature, produced by a number of different methods, including Abundance Matching \citep{Moster2010}, Empirical Modelling \citep{Moster2018, Behroozi2019}, Conditional Stellar Mass Functions \citep{Yang2012}, GGL
    \citep{Wang2021, Dvornik2020, zu2015, Hudson2014}, and a Milky Way Satellite Census \citep{Nadler2020}.
    Right: The posteriors for the three SHMR parameters of Eq.~\eqref{eq:shmr_model} for the joint fit to the measurements using  the three mass samples.
    The three satellite fraction constraints and those from the analysis of each mass sample individually can be found in Appendix~\ref{sec:SHMR_app}.
    }
    \label{fig:overplot_smhr}
\end{figure*}

The left-hand panel of Figure~\ref{fig:overplot_smhr} shows the $1\sigma$ posterior of the underlying mean stellar-to-halo mass relation constrained using the joint fit to the measurements from the three mass samples and its comparison to existing literature. The right-hand panel shows the posterior of the model parameters in the SHMR (Eq.~\eqref{eq:shmr_model}), from which the left-hand panel is derived.
In the next section, we discuss our results and constraints on the model and compare them to the literature. 

\section{Discussion}\label{sec:disc}

\subsection{Stellar Mass Halo Mass relation}
We find that the CDM profiles are good fits to the observed dwarf lensing profiles for each of the stellar mass bins. The galaxies that we observe lie in haloes with a median mass between $10^{10.6}-10^{11.4} \msun$.
We also independently fit NFW profiles to the lensing signal. The constraints are shown in Figure~\ref{fig:nfw_posteriors}. These fits are useful to explore as they do not assume an underlying $\Lambda$CDM universe a priori. We find that the NFW fits are consistent with our full simulation-based modelling overall. However, we note that the NFW fits prefer a high concentration for the lowest mass bin and a shallower concentration for the middle bin, shallower than what is expected from the concentration mass relation from $\Lambda$CDM.
The stellar halo mass relation constrained from this work probes a crucial region in the low-mass range between $10^{10}$ and $10^{12}\msun$, which has not been directly measured previously. Our SHMR parameters, particularly the slope $\alpha$, are best constrained when the three bins are fit jointly. We find a value of $\alpha \sim 2$ for these mass scales, which is largely consistent with literature \citep{Wechsler_2018}. Since we do not explicitly model the turnover in the SHMR at higher halo masses in this work, we check how our constraints are affected by using a prior on the intercept, i.e., the stellar mass, at $10^{12}$ halo mass that is derived from \citep{Behroozi2019}. We find that our values for the slope and the scatter are consistent with the full analysis. Note that most of the mean relations from the literature that are shown in Figure~\ref{fig:overplot_smhr} are extrapolations from measurements at higher masses. 
The only statistical constraint from data at low stellar masses is from \cite{Nadler2020}, which uses the counts of low-mass Milky Way satellite galaxies to constrain the SHMR down to the faintest limits of observed galaxies. Our results are statistically consistent with this work. Note that our stellar masses are complementary and probe the range between the existing low-mass and high-mass measurements.

\subsection{Satellite fraction}

\begin{figure}
    \centering
    \includegraphics[width=\columnwidth]{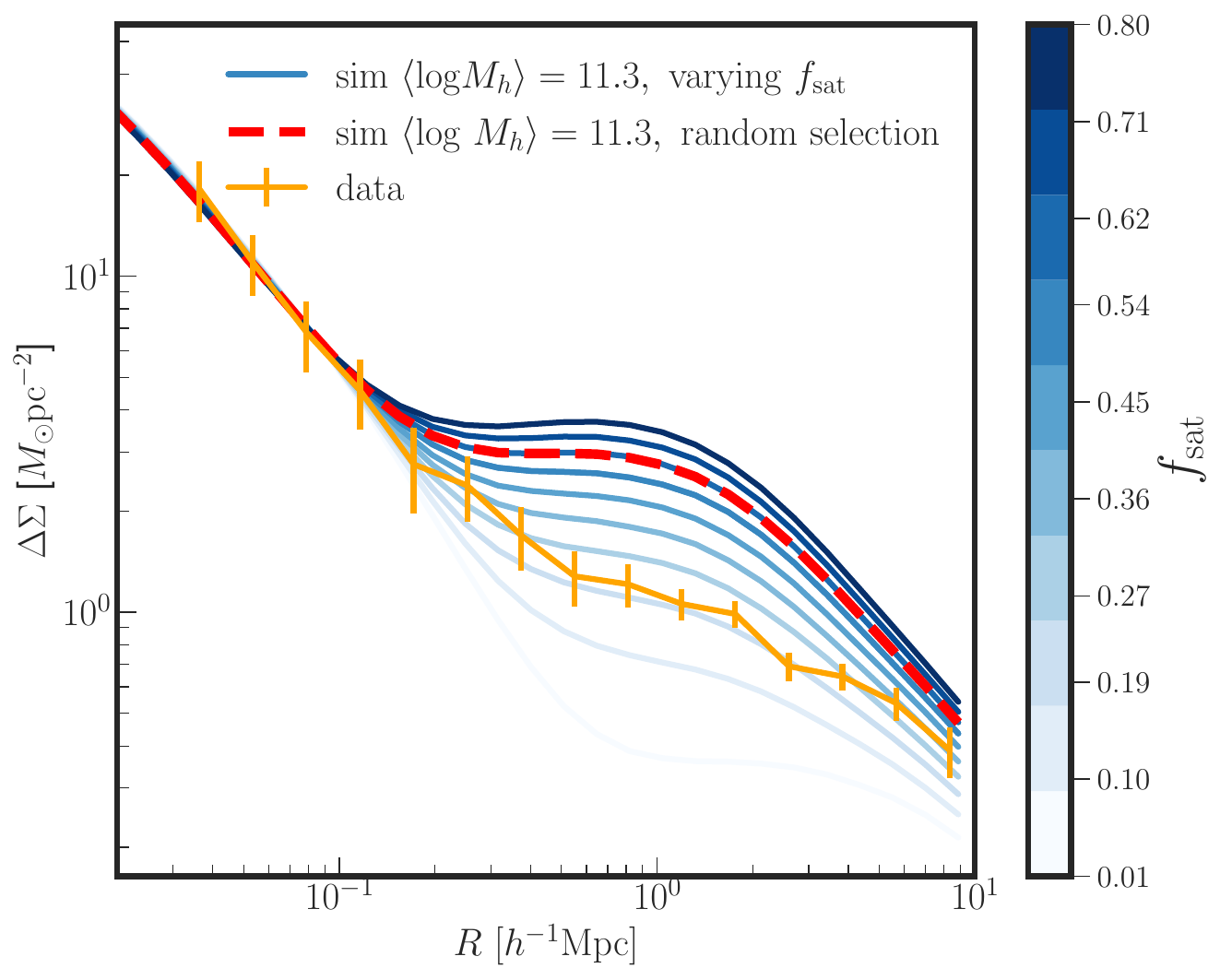}
    \caption{Stacked excess surface density profile predictions from the CDM N-body simulation for a \textsc{HIGH}-mass sample with mean peak halo mass of $10^{11.3} ~h^{-1}\msun$, and a varying fraction of subhaloes, $f_{\rm sat}$, contributing to the samples, indicated by the blue colour gradient (in bins between 0 to 0.8). The outer profile is strongly affected by the satellite fraction. The orange points are the \textsc{HIGH} mass sample $\Delta \Sigma(R)$ lensing measurements, which constrain a mean $f_{\rm sat}=0.219\pm0.022$. The red dashed line corresponds to the stacked $\Delta \Sigma(R)$ prediction for a random selection of haloes at the same mean mass. The corresponding intrinsic satellite fraction in a CDM universe is $f_{\rm sat}\sim0.6$.}
    \label{fig:sim_model}
\end{figure}

In this work, we find that the measured profiles across the full radial range require that a significant fraction of the galaxies in each of the three stellar mass bins live as satellites of more massive haloes, i.e., they are subhaloes. Our constraints indicate that close to $\sim 20\%$ of the galaxies in our lens samples are satellites. 

To further understand the implications of this result, Figure~\ref{fig:sim_model} demonstrates the effect of varying the satellite fraction on the measured stacked lensing profile of samples of haloes with the same mean $M_{\rm peak}$. 
The blue curves show the stacked excess surface density profiles of haloes at a narrow bin around a mean $\log (M_{\rm peak}/\msun)=11.3$, which corresponds to the mean mass of the \textsc{HIGH} mass sample.
The profile in the inner regions traces the usual NFW-like inner halo, however, as the satellite fraction is increased the density in the outer region increases systematically.
In particular, as $f_{\rm sat}$ is increased a shallow peak appears in the outer region, tracing approximately the off-centred profiles of host haloes (integrated over the mass function above the satellite mass--scale).
The location of the peak is at $\sim 1 \,h^{-1} {\rm Mpc} $, which is the characteristic scale of the virial radius of a cluster--mass halo, the largest hosts within which these satellites reside; beyond this scale, the excess density eventually falls off. 

Figure~\ref{fig:sim_model}  also shows the stacked $\Delta \Sigma$ profile for a random sampling of haloes (i.e., without adjusting $f_{\rm sat}$) at the given mean $M_{\rm peak}$ (red dashed curve), for a $\Lambda$CDM universe. We note by comparison to the blue curves that a random sampling contains a significant fraction of subhaloes, close to $\sim 60\%$. Note that this specific number depends on our exact definition of `satellites'; typically satellites are defined based on whether the halo is within the virial radius of a larger object (PID$=-1$ condition in \textsc{rockstar}), giving a satellite fraction closer to $20-30\%$ in the halo mass range shown in the figure \citep{2017ApJ...834...37L}. As expected the satellite fraction, using our definition, appears to be larger as splashback subhaloes are taken into account.   

This random selection model is also compared with the measured excess surface density profile for the \textsc{HIGH} stellar mass bin (orange curve). We note that the inferred satellite fraction from the observations for this sample is low compared to a random selection of galaxies, implying that we have a biased sample with fewer satellites than a random selection of galaxies in a $\Lambda$CDM universe. The other two stellar mass bins show the same behaviour. 
A likely explanation for this bias toward lower satellite fraction is that the \textit{candidate low-mass sample}, from which the lens samples were selected, was constructed using the same photometric selection as the SAGA Primary Targeting Region (Eqs.~\eqref{eq:colour_cut} and \eqref{eq:sb_cut}), which was optimised for completeness only for $z<0.015$ galaxies (\citealt{Mao_2021}, and see discussion of the redshift dependence of colour cuts in \citealt{Elise2023}). In other words, for $z>0.015$ galaxies, the colour cut we applied will preferentially select bluer galaxies at a fixed stellar mass. 
Since the redshift distributions of our three lens samples all peak at $z\ge0.08$ (Figure~\ref{fig:true_est_distrb_TOMO}), these lens samples are likely missing redder galaxies if compared to a truly stellar-mass selected sample. 

We note that our model for the satellite fraction does not include dependence on stellar mass or redshift; however, the satellite fraction bias introduced by the colour cut should vary with stellar mass and redshift. 
For these reasons, we take our constraints on the satellite fraction as a lower limit for an unbiased sample. 
Note that completeness is not essential for the lensing measurement itself, and the missing redder galaxies would not impact the mass constraints significantly. Our results simply imply that bluer galaxies at dwarf galaxy scales live in relatively low-density regions and appear to have a lower satellite fraction. This is consistent with models that invoke environmental quenching of star formation, as bluer, star-forming galaxies are found preferentially in low-density regions. While similar colour--environment relations are well known for brighter galaxies, there are fewer studies in this stellar mass regime.

\section{Summary and Outlook}\label{sec:summary}
While the $\Lambda$CDM model has been successful at fitting cosmological observations, the nature of dark matter remains unknown.
Weak lensing measurements of dwarf galaxies provide an avenue to test the properties of dark matter in two ways: by studying the excess surface mass density profiles, $\Delta\Sigma(R)$, and via constraints on the SHMR in the low-mass regime.
The aim of this work has been to explore a method of extracting large samples of dwarf-mass galaxies from photometric data in order to measure and model their galaxy--galaxy lensing mass profiles.
We demonstrate that this technique can create exciting opportunities for dwarf lensing with its application to upcoming surveys. The main results of this study are:
\vspace{-0.8em}
\begin{itemize}
    \item \textbf{A new galaxy selection technique that incorporates the tandem use of photometric and spectroscopic surveys to extract a statistical sample of low-mass objects.}
First, we obtained a \textit{spectroscopic calibration sample} from the SAGA Survey that is preferentially low-redshift and low-mass. The photometric selection of this sample was also applied to DES DR2 objects to create a \textit{candidate low-mass sample}. A self-organising map (SOM) is trained with the \textit{candidate low-mass sample}, and each SOM `cell' is labelled with the redshifts and stellar masses of the \textit{spectroscopic calibration sample}. Using this framework, we extract further subsets of the data according to the average stellar mass of the SOM cell.
    \item \textbf{We produce three \textit{low-mass samples}, each with $\mathbf{>140,000}$ galaxies and with calibrated stellar mass distributions.} The median stellar masses of the three samples are $\log_{10}(M_*/\msun) = [8.52\substack{+0.57 \\-0.76}, 9.02\substack{+0.50 \\ -0.64}, 9.49\substack{+0.50 \\ -0.58}]$. 
    \item  \textbf{We measure the stacked excess surface mass density profiles of the \textit{low-mass samples}} using the background DES Y3 lensing data. The signal produced for the lowest mass sample, with $\langle\log_{10}(M_*/\msun)\rangle< 9$, has a S/N $>13$, providing encouraging prospects for this method using more statistically powerful spectroscopic and photometric surveys.
    \item \textbf{ Using a simulation-based forward-modelling approach, we fit the measurements using a power-law model for the SHMR} while varying the satellite fraction of the sample
    . We directly constrain the SHMR with weak lensing for the first time in this low-mass regime. The SHMR constrained by a joint fit to the three mass sample measurements is consistent with model predictions in the literature. 
    \item \textbf{The satellite fraction is tightly constrained at values distinct from that predicted by a mass-limited sample in CDM} for all three signals. This suggests that either we deviate from $\Lambda$CDM within this mass range, or, more likely, that our sample exhibits a selection bias, preferentially selecting central galaxies in our data. 
    \item \textbf{We estimate the average halo mass for each \textit{low-mass sample} using both simulation-based modelling and a simple NFW profile} fitted out to the approximate virial radius of our largest galaxies' dark matter haloes. The median masses obtained using the two approaches are consistent. 
\end{itemize}

\textbf{Outlook:} In the absence of complete spectroscopic coverage of photometric surveys, selecting large galaxy samples galaxies based on their stellar mass is challenging, particularly in the dwarf galaxy regime.
The SOM approach allows us to leverage much smaller, information-rich, spectroscopic surveys to access dwarf samples over the entire footprint of photometric surveys. 
Our proof-of-concept analysis has highlighted two avenues that would improve the robustness and the constraining power of future work: (1) increased signal-to-noise of the weak lensing measurements, and (2) better-characterised galaxy samples, which would allow for better separation of galaxies by stellar mass. 

The first point will be achieved either with a more statistically powerful background source sample or a larger foreground lens sample. Ongoing surveys, such as Hyper Suprime-Cam \citep[HSC;][]{HSC2017} and next-generation surveys such as Rubin LSST \citep{Rubin_2019} and Euclid \citep{euclid2011} will do this naturally, with unprecedented deep large area photometric data. Deeper imaging data will also afford the opportunity to detect and extract fainter objects, and therefore to probe smaller halo masses.  For example, with the Rubin LSST 10-year dataset, it is possible that we can measure the dark matter profiles of dwarf galaxies living in haloes of $M_\text{halo} \sim 10^{9-10} \, \msun$ \citep{1902.01055}.

Second, a better-characterised dwarf sample can be achieved with additional spectra over a broader selection. For example, extending to fainter magnitudes than $r_o=20.75$ would enable an increase in usable photometry for defining the dwarf sample. 
A wider spectroscopic sample would also help to study and mitigate selection effects in the constructed dwarf sample by allowing for constraints on the satellite fraction and SHMR in the absence of the initial \textit{candidate low-mass sample} selection criteria. Even with the same selection, increasing the size of the spectroscopic sample would allow for increased SOM resolution without compromising a well-defined stellar mass distribution within each cell, and would allow the use of additional galaxy properties in the SOM.  We expect that both of these would improve the characterisation of the galaxy population and enable splitting the data into finer stellar mass bins. 

While the SAGA Survey has laid the groundwork for expanding low-redshift spectroscopic samples, the Dark Energy Spectroscopic Instrument Survey \citep[DESI;][]{DESI2014} will quickly increase the available spectroscopic data in this low-redshift and low-mass regime through the DESI Bright Galaxy Survey \citep{BGS} and the DESI LOW-Z Secondary Target Survey \citep{Elise2023}. The DESI LOW-Z program was built upon the SAGA strategy to collate a complete sample of targets with $z \sim 0.03$. It obtained redshifts for more than 22,000 dwarf $(M_* < 10^9 \msun)$ galaxies in the early months of the DESI survey and is continuing to operate as a low-priority DESI program. Looking ahead, the Prime Focus Spectrograph \citep{Takada2014}, Euclid \citep{euclid2011}, 4-metre Multi-Object Spectroscopic Telescope \citep{4MOST}, and future generations of DESI provide an opportunity to select large, well-understood dwarf galaxy samples.

This work demonstrates the effectiveness of the tandem use of photometric and spectroscopic surveys to extract a statistical sample of dwarf galaxies, particularly suitable for weak lensing. The outlook of this technique is exciting as the quality, quantity, and depth of both spectroscopic and imaging data continue to increase. These data will provide the opportunity to divide galaxy samples into finer mass bins, probe to smaller halo masses, and distinguish the impact of various galaxy properties on the inner regions of haloes. In the future, this methodology will allow for a new probe of the galaxy--halo connection, the impact of baryonic feedback on halo profiles, and novel tests of dark matter models. 

\section*{Acknowledgements}
The authors would like to thank Debora Sijacki and Angus Wright for useful discussions in the early stages of this work. Alexandra Amon is supported by a Kavli Fellowship at KICC. 
This work received partial support from the Kavli Institute of Particle Astrophysics and Cosmology at Stanford University.
Part of this work was performed at the Aspen Center for Physics, which is supported by National Science Foundation grant PHY-1607611.This project used the PARAM Brahma supercomputing facility at IISER Pune, which is part of the National Supercomputing Mission of the Government of India.

The SAGA Survey (sagasurvey.org) is a spectroscopic survey with data obtained from the Anglo-Australian Telescope, the MMT Observatory, and the Hale Telescope at Palomar Observatory. The SAGA Survey made use of public imaging data from the Sloan Digital Sky Survey (SDSS), the DESI Legacy Imaging Surveys, and the Dark Energy Survey, and also public redshift catalogues from SDSS, GAMA, WiggleZ, 2dF, OzDES, 6dF, 2dFLenS, and LCRS. The SAGA Survey was supported was supported by NSF collaborative grants AST-1517148 and AST-1517422 and by Heising–Simons Foundation grant 2019-1402.

This project has used public archival data from the Dark Energy Survey. Funding for the DES Projects has been provided by the U.S. Department of Energy, the U.S. National Science Foundation, the Ministry of Science and Education of Spain, the Science and Technology Facilities Council of the United Kingdom, the Higher Education Funding Council for England, the National Center for Supercomputing Applications at the University of Illinois at Urbana-Champaign, the Kavli Institute of Cosmological Physics at the University of Chicago, the Center for Cosmology and Astro-Particle Physics at the Ohio State University, the Mitchell Institute for Fundamental Physics and Astronomy at Texas A\&M University, Financiadora de Estudos e Projetos, Funda{\c c}{\~a}o Carlos Chagas Filho de Amparo {\`a} Pesquisa do Estado do Rio de Janeiro, Conselho Nacional de Desenvolvimento Cient{\'i}fico e Tecnol{\'o}gico and the Minist{\'e}rio da Ci{\^e}ncia, Tecnologia e Inova{\c c}{\~a}o, the Deutsche Forschungsgemeinschaft, and the Collaborating Institutions in the Dark Energy Survey.
The Collaborating Institutions are Argonne National Laboratory, the University of California at Santa Cruz, the University of Cambridge, Centro de Investigaciones Energ{\'e}ticas, Medioambientales y Tecnol{\'o}gicas-Madrid, the University of Chicago, University College London, the DES-Brazil Consortium, the University of Edinburgh, the Eidgen{\"o}ssische Technische Hochschule (ETH) Z{\"u}rich,  Fermi National Accelerator Laboratory, the University of Illinois at Urbana-Champaign, the Institut de Ci{\`e}ncies de l'Espai (IEEC/CSIC), the Institut de F{\'i}sica d'Altes Energies, Lawrence Berkeley National Laboratory, the Ludwig-Maximilians Universit{\"a}t M{\"u}nchen and the associated Excellence Cluster Universe, the University of Michigan, the National Optical Astronomy Observatory, the University of Nottingham, The Ohio State University, the OzDES Membership Consortium, the University of Pennsylvania, the University of Portsmouth, SLAC National Accelerator Laboratory, Stanford University, the University of Sussex, and Texas A\&M University. Based in part on observations at Cerro Tololo Inter-American Observatory, National Optical Astronomy Observatory, which is operated by the Association of Universities for Research in Astronomy (AURA) under a cooperative agreement with the National Science Foundation. 

\section*{Data availability}

\bibliographystyle{mnras}
\bibliography{bib.bib}

\appendix
\section{Sample selection}

\subsection{Consistency between spectroscopic calibration sample and the candidate low-mass sample}\label{sec:repro_app}

In this Appendix, we comment on the inherent differences between the \textit{spectroscopic calibration sample}, and \textit{candidate low-mass sample} despite their identical photometric selection criteria. These two samples might differ for three reasons:

 1) The \textit{spectroscopic calibration sample} does not encompass all photometric objects in the photometric space; hence, the SOM may introduce a bias in the estimated mass distributions. Our \textit{spectroscopic calibration sample} is highly complete as it comes from the SAGA Survey, which has a $>$90\% redshift coverage in the SAGA Primary Targeting Region ($>$85\% in the $18 < r_o < 20.75$ region). In addition, if the remaining incompleteness correlates with the parameters used to train the SOM, this selection bias will be mitigated by the SOM selection. 2) While both the SAGA imaging and DES imaging use the Dark Energy Camera, the processing from SAGA imaging comes from DESI Legacy Imaging Surveys. 3) Finally, there is noise in the two samples. 

\begin{figure}
    \centering
    \includegraphics[width=\columnwidth]{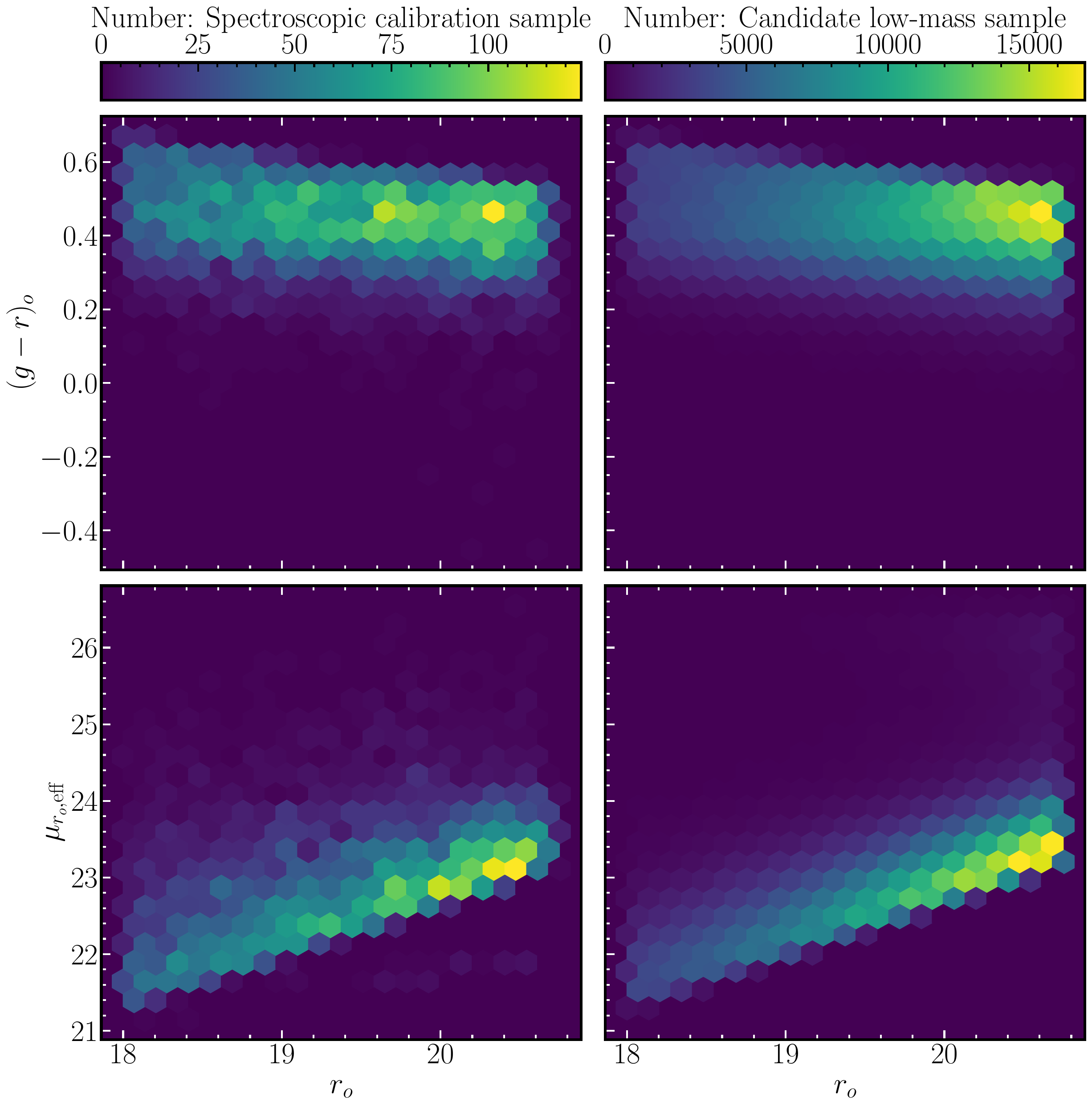}
    \caption{A comparison between the density of points in the different selection planes (rows). Columns 1 and 2 show density in the \textit{spectroscopic calibration sample} and the \textit{candidate low-mass sample} respectively.
    Rows from top to bottom show colour--magnitude, surface magnitude--magnitude.
    This plot was used as a visual check for the representative nature of the \textit{spectroscopic calibration sample} distributions compared to the \textit{candidate low-mass sample} distributions.}
    \label{fig:2d_compare}
\end{figure}

We visually compare the galaxy property distributions in the \textit{spectroscopic calibration sample} to the \textit{candidate low-mass sample} in Figure~\ref{fig:2d_compare}. Specifically, we compare the data in the 2D planes of colour--magnitude and colour--surface brightness for the noisier calibration data on the left and the candidate objects on the right. We see broad similarities between the two distributions. At faint magnitudes, we see some differences in the colour distributions, but these appear to be within the noise. Nevertheless, the SOM characterises the data into narrow cells using these three properties, so any underlying differences in the samples are mitigated in the limit of a sufficiently resolved SOM. 
In addition, we get further indication that our \textit{spectroscopic calibration sample} is similar to our main \textit{candidate low-mass sample} by the similarity between the filled and dashed distributions of Figure~\ref{fig:dual_TOMO_and_BLC}. That is, after the SOM selection, the distributions within each mass bin are well described by the similarly assigned calibration objects without any reweighting.

\subsection{SOM training}\label{sec:SOM_app}

\begin{figure*}
    \includegraphics[width=\textwidth]{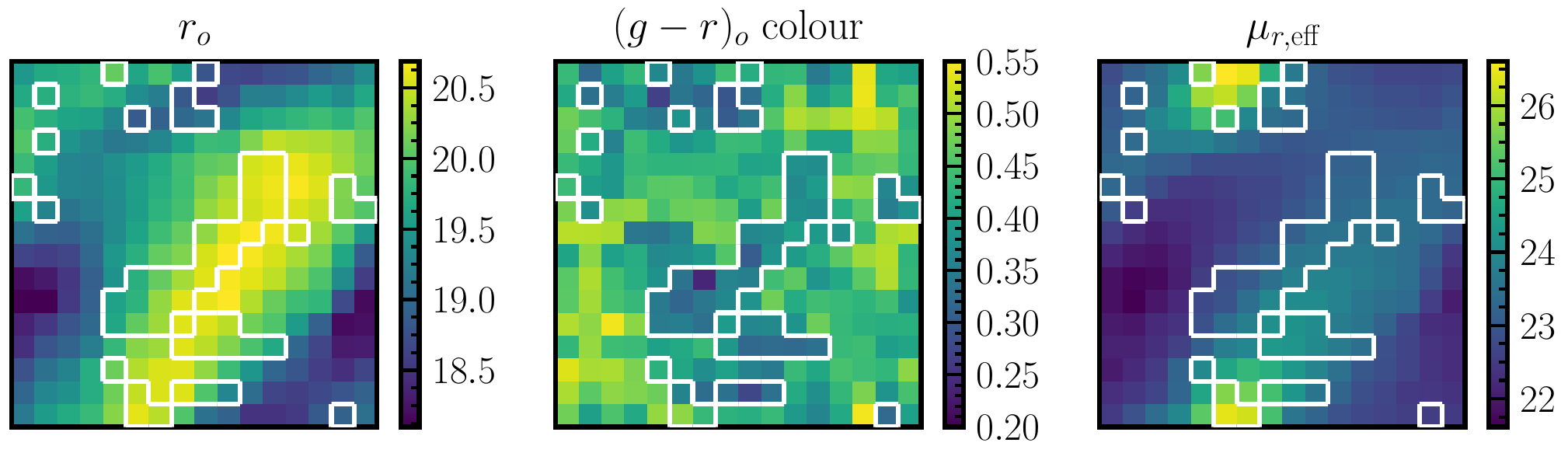}
    \caption{The SOM after training on the candidate low-mass sample, after selection processes outlined in Section~\ref{sec:data} and Table~\ref{tab:cut-fractions}.
    Each panel displays the final state of the SOM in one of the three training properties.
    The colour bar indicates the value assigned to each cell of the SOM in a given property.
    These three training properties are shared by both the photometric and spectroscopic datasets.
    The cells outlined in white are those that have $\langle \log_{10}(M_*/\msun)\rangle < 8.75 $, and are thus assigned to the lowest mass bin of our analysis.}
    \label{fig:trained_SOM}
\end{figure*}

\begin{figure*}
    \centering
    \includegraphics[width=\textwidth]{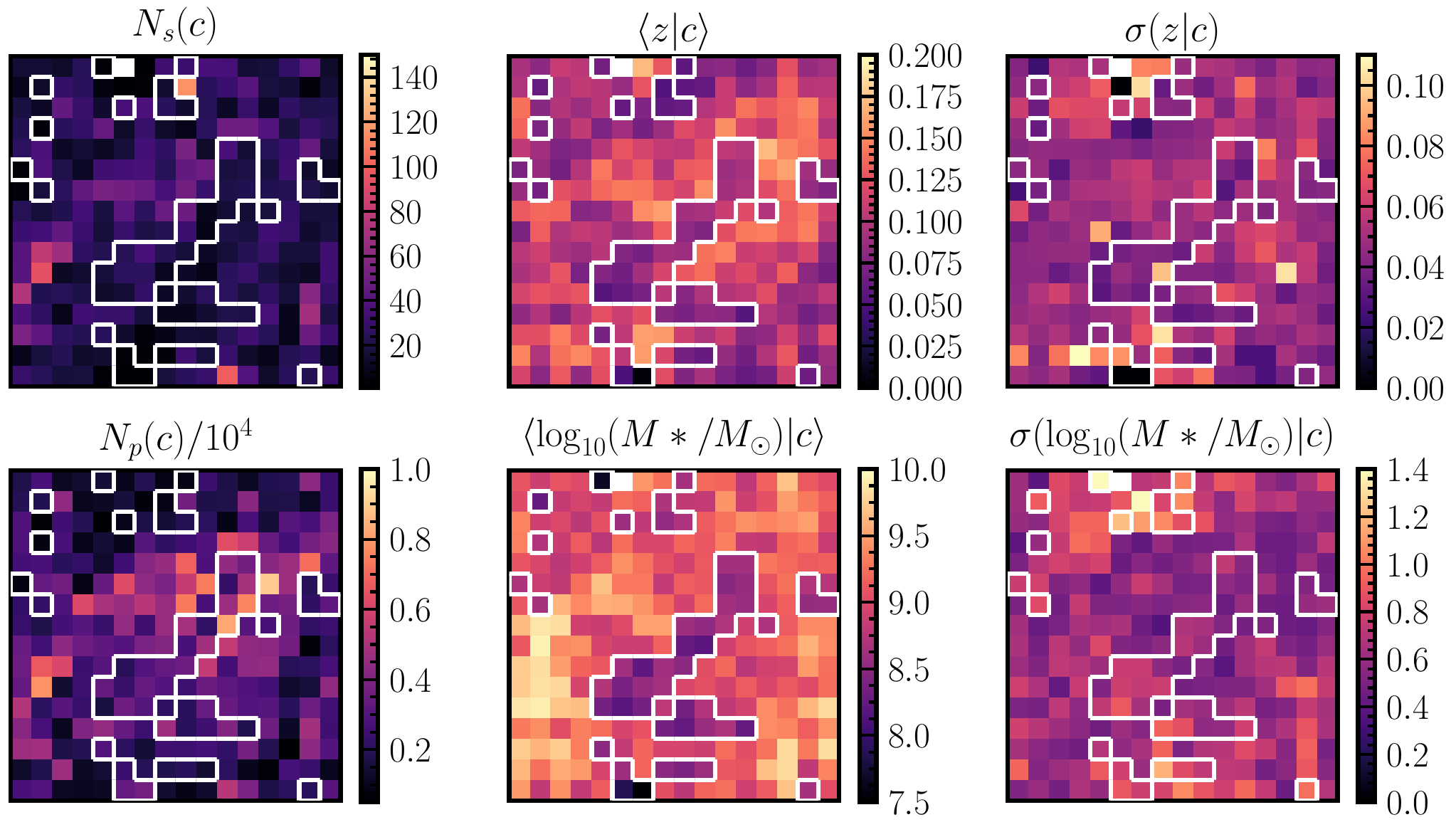}
    \caption{SOM statistics: The number of assigned \textit{spectroscopic calibration sample}, $N_{\rm s}$, and \textit{candidate low-mass sample}, $N_{\rm p}$, for each cell is shown in left panel. The middle panel is showing the average inferred redshift and stellar mass from the \textit{spectroscopic calibration sample} for each cell (middle), and the right displays the $1\sigma$ spread on each quantity for a given cell.
    As in Figure~\ref{fig:trained_SOM}, the cells outlined in white are those that have $\langle \log_{10}(M_*/\msun)\rangle < 8.75 $, and are thus assigned to the lowest mass bin of our analysis.
    Comparison with the mass panel clearly indicates how the low-mass cells are selected.
    In each of these panels (bar $N_p(c)/10^4$) a white cell indicates that no objects from the \textit{spectroscopic calibration sample} are assigned to that cell.}
    \label{fig:assigned_SOM}
\end{figure*}

Here we discuss a brief overview of the training process for the SOM, the method and naming conventions follow those outlined in \cite{Buchs_19}, which we refer the reader to for more detail.
We define our SOM as a $16\times16$ array of cells, $\mathcal{C}$, such that the total number of cells is given by $C=256$.
The total number of cells is limited by the abundance of spectroscopic data, $C=256$ was chosen to ensure a mean calibration occupation of $\sim 25$, to improve confidence in the inferred distributions.
The dimensionality of our training data is $m=3$, but can be trivially extended when applied to larger spectroscopic datasets.
Each cell, $c\in \mathcal{C}$ is initialised with a random weight vector $\mathbf{w}_k\in\mathbb{R}^m$, where $k=1, \cdots, C$.
The SOM is trained on a subset of the \textit{candidate low-mass sample}, $\mathcal{S}$, of length $S=200,000$, with each element $s\in\mathcal{S}$ having a corresponding vector $\mathbf{x}_{\rm s}\in\mathbb{R}^m$, where $s=1, \cdots, S$.
Each step of the training process picks a random element $s\in\mathcal{S}$ and the $\chi^2$ distance is calculated between $s$ and all cells $c\in\mathcal{C}$ following
\begin{equation}
	\Delta\chi^2 = (\mathbf{x}_{\rm s} - \mathbf{w}_c)\Sigma^{-1}(\mathbf{x}_{\rm s} - \mathbf{w}_c)^\intercal,\label{eq:chi2}
\end{equation}
where $\Sigma$ is the covariance matrix for our system, with a diagonal $\sigma_1,...,\sigma_m$.

If $t$ is the current time step of the training, the weights are updated as:
\begin{equation}
	\mathbf{w}_k(t+1) = \mathbf{w}_k(t) + a(t) H_{b, k}(t)[\mathbf{x}_{\rm s}(t)- \mathbf{w}_k(t)],
\end{equation}
where $a(t)$ is the learning rate function, determining the amount of change a given datum causes to the SOM, and $H_{b,k}(t)$ is the neighbourhood function, determining the change of weight vectors for cells near to the best-fitting cell.
The specifics of these functions are discussed in \cite{Buchs_19}.

We train on the \textit{candidate low-mass sample} so that any of its objects that do not closely match those of the \textit{spectroscopic calibration sample} will be placed in cells that do not acquire a calibration label.
Conversely, if the SOM were trained on calibration data all candidate low-mass objects would be allocated to a mass distribution, even if the match were poor.
Examples of cells that benefit from this training are indicated in solid white in Figure~\ref{fig:assigned_SOM}, and are not assigned to a stellar mass bin.

Figure~\ref{fig:trained_SOM} shows the final results of training the SOM.
The left two maps displaying the density of objects per cell for the different catalogues is a useful check of the representative nature of the calibration sample and the candidate low-mass sample, with dark and lighter regions falling in similar clumps throughout the map.
The right-hand two maps are a useful indicator of low spread in redshift and stellar mass for given cells, suggesting that for a given cell the mass and redshift estimates are fairly well defined.
The two middle maps contain the assigned quantities, with the stellar masses being the results we are primarily interested in, only needing the redshift distributions for our calculation of $\Delta\Sigma$.
Spatial clustering in both is another good indicator that once the mass bins have been extracted from the SOM, the objects contained within will all be similar.
If the maps of our assigned quantities did not display prominent clustering the validity of this method would be questionable, as it would not be reasonable to assume a relationship between the measured quantities and the stellar masses.

\section{Validating the lensing signal}\label{sec:lens_app}
\begin{figure*}
    \centering
    \includegraphics[width=\textwidth]{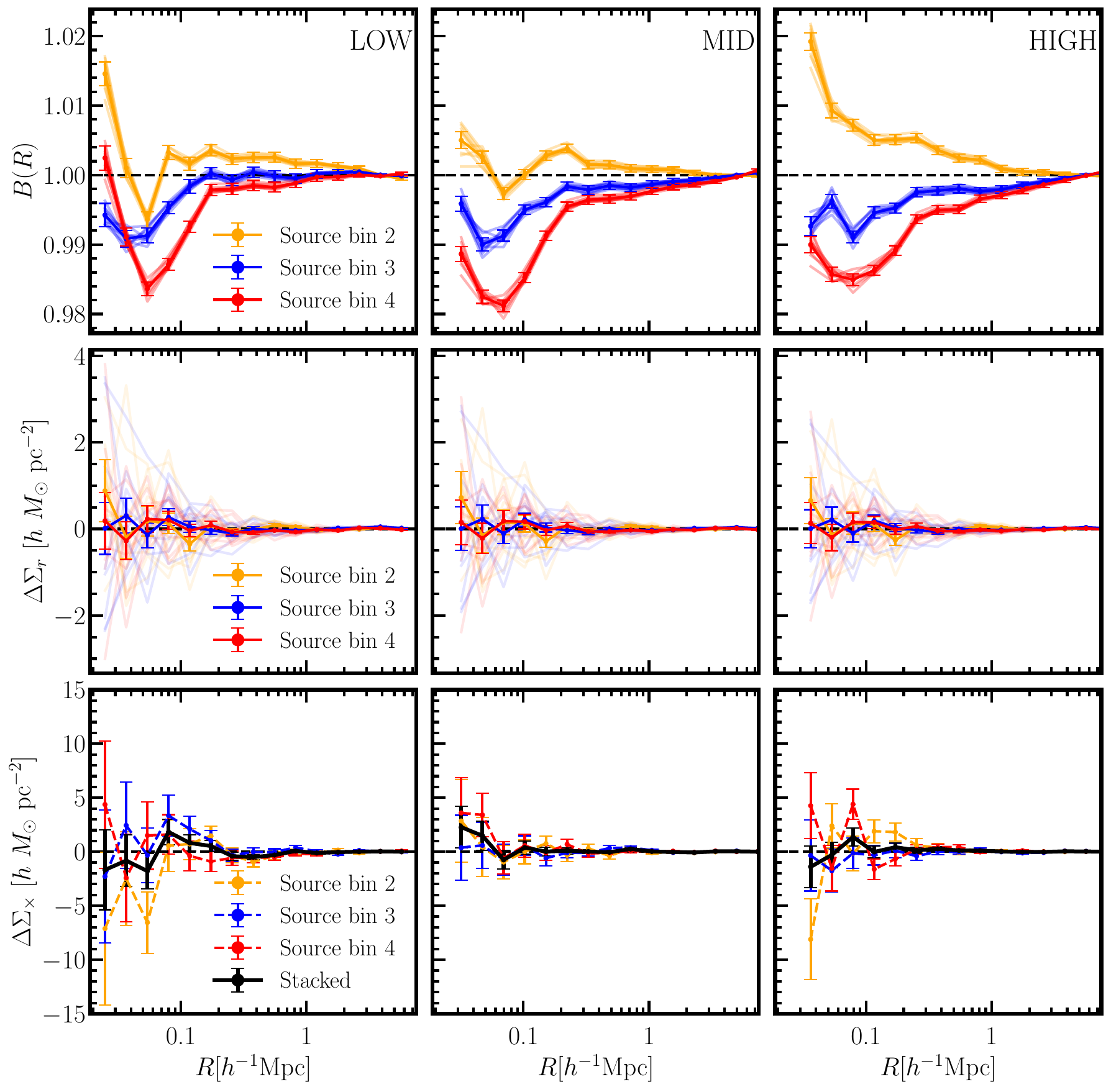}
    \caption{Diagnostic plot for the lensing measurements using the \textsc{LOW}, \textsc{MID} and \textsc{HIGH} \textit{low mass samples}, from left to right.
    From top to bottom the rows show the boost factor (see section~\ref{sec:est}), the $\Delta\Sigma$ around random points, and the cross shear, converted to the $\Delta\Sigma$ level for clarity.
    In each case, the yellow, blue, and red indicate source bins 2, 3, and 4 respectively, with the black being the stacked results.} 
    \label{fig:mega_diagnostic}
\end{figure*}

\subsection{Cross shear null test}
As discussed in Section~\ref{sec:est}, GGL only induces a tangential shear, however, an alternative shear, offset by $45^\circ$, can also be measured around our lens sample.
As the entirety of GGL-induced shear is expected to be in the tangential component, we can use measurements of the `cross-shear' to check for systematic effects and to validate the implementation of our pipeline.
We measure the cross-shear for each of our mass bins, the results from which are shown in Figure~\ref{fig:mega_diagnostic}, visually confirming the cross-shear to be consistent with zero.
We also compared it to a null signal and found all stacked cross-shears presented $\chi^2_{\nu, \mathrm{null}} < 1$, again indicating no statistically significant signal.

\subsection{Random Catalogues}

For both the boost and random signal described in Section~\ref{sec:est}, we require a catalogue of random lens positions within the DES Y3 footprint, and with the same redshift distributions as our lenses. 
We stack the results from 10 random catalogues, each with 1.2 million positions so that the total number of random objects is at least 30 times the size of our lens catalogue.
Each catalogue consists of a spatially uniform distribution of points inside the DES Y3 footprint and is given the same underlying redshift distribution as the relevant mass bin.
The calculation is performed with the same random sample for each source bin.
Results of the random measurement are displayed in the middle row of Figure~\ref{fig:mega_diagnostic}, all being far smaller than respective lens signals, as expected.

We use the same random catalogues to calculate the boost factor, $B(R)$, the results from which are shown in the top row of Figure~\ref{fig:mega_diagnostic}.
We can see as we head further from the position of our lens the boost factor of all bins tends to zero.
This is expected as we expect to tend towards a uniform distribution the further we are away from our lenses.
The boost factors measured in this work are smaller than those calculated in previous GGL work \citep{Voice2022}, which we attribute to our low redshift lens sample compared to the three DES Y3 source bins used.
Additionally, there appears to be some effect producing a negative boost (an indication of underdensity) near the lenses, which may only be observable when dealing with such well-separated lens and source distributions, hence why only source bin 2 produces a positive boost.

\section{Modelling}

\subsection{{\texorpdfstring{SHMR and f\textsubscript{sat} posteriors}{SHMR and fsat posteriors}}}\label{sec:SHMR_app}

Here, we summarise the posterior of the simulation-based inference using the SHMR and $f_\mathrm{sat}$ model. For all parameters, we imposed uniform priors, which are reported in Table~\ref{tab:SHMR_params}.
The median value, with $16^\mathrm{th}$ and $84^\mathrm{th}$ percentile bands, of the marginalised posteriors are also summarised for each parameter.

\begin{figure}
    \centering
    \includegraphics[width=\columnwidth]{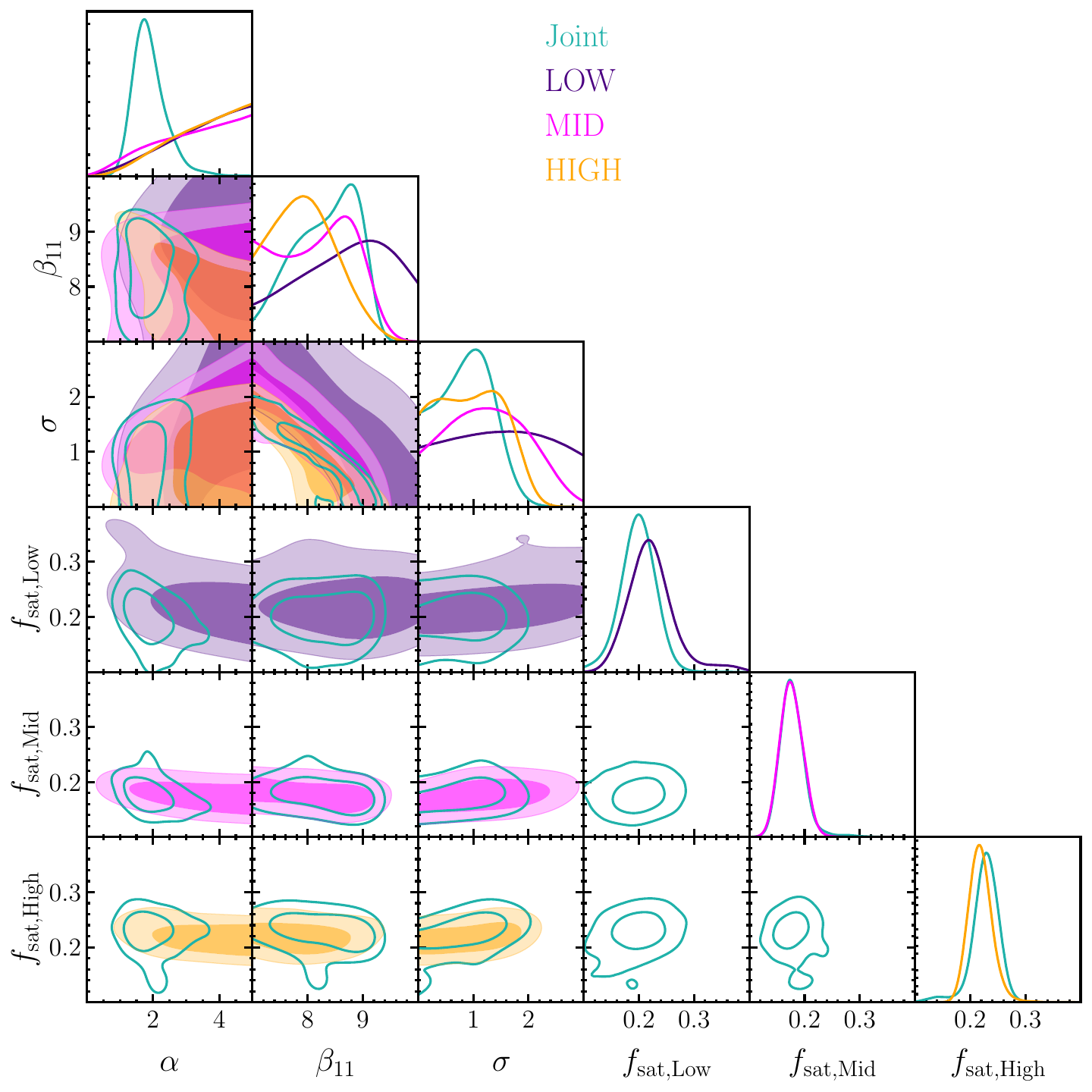}
    \caption{Full posterior for simulation-based model fits, for the \textsc{LOW}, \textsc{MID}, and \textsc{HIGH} mass samples in purple, pink, and orange, and the joint fit to all three mass samples overlaid in teal (a subset of which are shown in the right hand of Figure~\ref{fig:overplot_smhr}). The parameters shown are the  slope, $\alpha$, the intercept about a halo of mass of $10^{11} \msun$, $\beta_{11}$ and scatter, $\sigma$ in the SHMR, and the satellite fraction $f_\mathrm{sat}$. }
    \label{fig:shmr_posteriors}
\end{figure}

The full posterior is shown in Figure~\ref{fig:shmr_posteriors} for the joint constraint and for each of the \textsc{LOW}, \textsc{MID}, and \textsc{HIGH} mass bins. The constraints on the slope and intercept are difficult to interpret from the analyses using individual \textit{low-mass samples}. We find that the scatter increases for lower masses, although the constraint at lower masses is less precise. The satellite fractions show little degeneracy with the SHMR parameters.

\begin{table}
    \centering
    \def\arraystretch{1.5}%
    \caption{Summary of the uniform priors and median values of the marginalised posteriors for the SHMR and $f_\mathrm{sat}$ model parameters, described in Section~\ref{sec:model}. 
    The uncertainties quoted for the posteriors indicate the $16^\mathrm{th}$ to $84^\mathrm{th}$ percentile bands of the marginalised posteriors distributions. Quoted are the constraints when the model fits each mass bin separately and from an SHMR jointly constrained over all three bins.
    }
    \begin{tabular}{c|cccc}
        \hline
       Param & $\alpha$ & $\beta_{11}$ & $\sigma$ & $f_\mathrm{sat}$ \\
        \hline
        Prior &[0,5] & [7,10] & [0,3] & [0,0.8] \\ 
         \hline
        \textsc{LOW}  & $3.60\substack{+0.97\\ -1.43}$ & $8.76\substack{+0.76 \\ -1.00}$& $1.51\substack{+0.92 \\ -0.98}$& $0.22\substack{+0.04 \\ -0.04}$\\
        \textsc{MID} & $3.35\substack{+1.18 \\ -1.60}$& $8.26\substack{+0.65 \\ -0.88}$& $1.26\substack{+0.77 \\ -0.74}$& $0.18\substack{+0.02 \\ -0.02}$\\
        \textsc{HIGH} & $3.59\substack{+0.98 \\ -1.37}$& $7.95\substack{+0.61 \\ -0.58}$& $1.01\substack{+0.62 \\ -0.67}$& $0.22\substack{+0.02 \\ -0.02}$\\
        Joint (\textsc{LOW}) & $1.82\substack{+0.56 \\ -0.40}$& $8.41\substack{+0.54 \\ -0.72}$& $0.90\substack{+0.45 \\ -0.57}$& $0.20\substack{+0.03 \\ -0.03}$\\
        Joint (\textsc{MID})&&&&$0.18\substack{+0.02 \\ -0.02}$ \\
        Joint (\textsc{HIGH}) &&&&$0.23\substack{+0.02 \\ -0.02}$\\
        \hline
    \end{tabular}
    \label{tab:SHMR_params}
\end{table}

\subsection{Comparison to NFW}\label{sect:NFW}

\begin{figure}
    \centering
    \includegraphics[width=0.9\columnwidth]{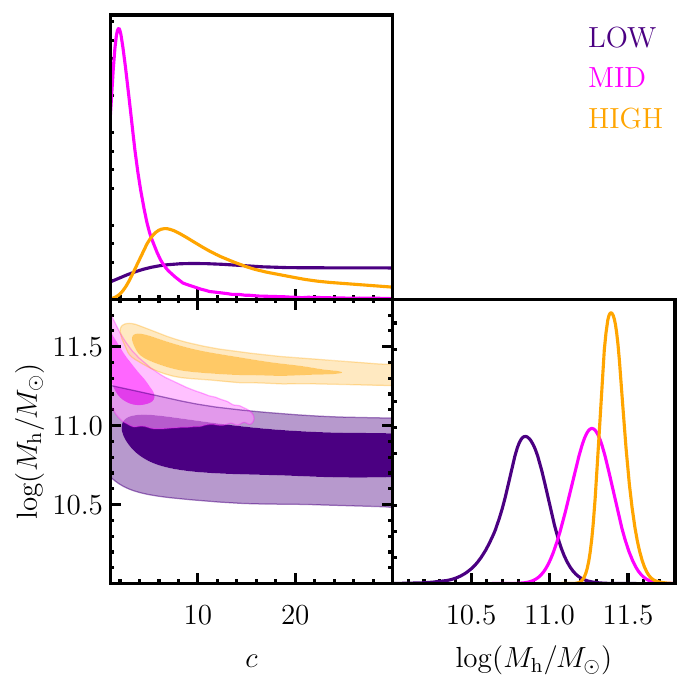}
    \caption{Posterior distribution for the NFW model fits to the measurements from the \textsc{LOW}, \textsc{MID}, and \textsc{HIGH} mass samples, shown in purple, pink, and orange respectively. The parameters shown are the halo mass, $\log_{10}(M_{\rm halo}/\msun)$, and the concentration, $c$.
    The best-fitting and median values of the concentration and mass can be found in Table~\ref{tab:NFW_results}.
    }
    \label{fig:nfw_posteriors}
\end{figure}

As a cursory test of both the results displayed in Figure~\ref{fig:dual_TOMO_and_BLC}, and the model fits of Section~\ref{sec:Modelling}, we fit a standard NFW profile \citep{NFW} to the inner regions of each signal.
We use the same mass convention as \textsc{SMDPL}, with a $200\times\rho_\mathrm{crit}$ overdensity criterion, using the $r_{200}$ radius.
The NFW profile has the functional form:
\begin{equation}
    \rho(r)=\frac{\delta_c \rho_{\rm crit}}{(r/r_{\rm s})(1+r/r_{\rm s})^2}\, ,
\end{equation}
where $r_{\rm s}\equiv r_{200}/c$ is the scale radius, and $\rm{c}$ is the concentration parameter, which defines a characteristic density, $\delta_c$, as
\begin{equation}
    \delta_c=\frac{200}{3}\frac{c^3}{\ln(1+c)-c/(1+c)}\, .
\end{equation}
The profile is specified with two free parameters. We fit the measurements to $\Delta\Sigma$ and  constrain the concentration and halo mass, with
\begin{equation}
    M = 4\pi \delta_c \rho_{\rm crit} r_{\rm s}^3 \left[\ln(1+c)-c/(1+c)\right]\, .
\end{equation}
The posterior is shown in Figure~\ref{fig:nfw_posteriors} for each \textit{low mass sample}. It is interesting that for all three bins, the masses are well--constrained to be consistent with the results of the SHMR--$f_\mathrm{sat}$ model.  In the \textsc{MID} mass bin, the concentration is well constrained. For this mass sample, the fit to the measurements deviates from the simulation-based profile in the innermost region, as shown in Figure~\ref{fig:final_fits}.

The corresponding values for the goodness of fit $\chi^2_\nu=\chi^2_\mathrm{min}/(N_{\rm d}-N_{\rm p})$, defined in terms of the number of data points, $N_{\rm d}$, and the number of parameters, $N_{\rm p}$, are quoted in Table~\ref{tab:NFW_results}. The fits are acceptable, although as we use a limited model and only the innermost data points, this model serves only as a cross-check.

\begin{table}
    \centering
    \caption{Summary of the posteriors for the NFW model halo mass and concentration, fitting the measurements of the \textit{low-mass samples}.
    Statistical errors quoted are the 84th and 16th percentiles of the posteriors.
    }
    \def\arraystretch{1.5}%
    \begin{tabular}{c|cc|ccc}
        \hline
        Sample &  \multicolumn{2}{c}{$\log_{10}(M_{\rm halo}/\msun)$} & \multicolumn{2}{c}{Concentration, $c$}  & $\chi^2_\nu$\\
        & Best Fit & Median & Best Fit & Median \\
        \hline
        Low & $10.79$ & $10.84\substack{+0.14 \\ -0.15}$ & 33.1 & $15.4\substack{+9.9 \\ -9.0}$ & 0.2 \\
        Mid & $11.32$ & $11.27\substack{+0.13 \\ -0.14}$ & 2.33 & $2.9\substack{+3.1 \\ -1.3}$ & 4.1\\
        High & $11.44$ & $11.40\substack{+0.08 \\ -0.07}$ & 7.94 & $10.7\substack{+9.8 \\ -5.0}$ & 0.2\\
        \hline
    \end{tabular}
    \label{tab:NFW_results}
\end{table}

\end{document}